\documentclass[journal]{IEEEtran}
\usepackage{amsmath,amsfonts}
\usepackage{algorithmic}
\usepackage{algorithm}
\usepackage{array}
\usepackage{textcomp}
\usepackage{stfloats}
\usepackage{url}
\usepackage{verbatim}
\usepackage{graphicx}
\usepackage{cite}

\usepackage{hyperref}
\usepackage{makecell}
\usepackage{bm}
\usepackage{amssymb}
\usepackage{subfigure}
\usepackage{amsthm}
\usepackage{multirow}
\usepackage{multicol}
\usepackage{float}

\newtheorem{proposition}{Proposition}

\begin{document}

\title{Multidimensional Graph Neural Networks for Wireless Communications}

\author{Shengjie Liu, 
	Jia Guo,~\IEEEmembership{Graduate Student Member,~IEEE,}
	and Chenyang Yang,~\IEEEmembership{Senior Member,~IEEE,}
\thanks{The codes are open-sourced on \href{https://github.com/LSJ-BUAA/MDGNN}{https://github.com/LSJ-BUAA/MDGNN}. This work was supported in part by the National Natural Science Foundation of China (NSFC) under Grant 62271024 and the National Key R\&D Program of China under Grant 2022YFB2902002. An earlier version of this paper was presented in part at the IEEE GLOBECOM 2022 [DOI: 10.1109/GLOBECOM48099.2022.10000790].}
\thanks{The authors are with the School of Electronics and Information Engineering, Beihang University, Beijing 100191, China (e-mail: liushengjie@buaa.edu.cn; guojia@buaa.edu.cn;  cyyang@buaa.edu.cn).}}

\markboth{IEEE Transactions on Wireless Communications, Accepted, 2023}%
{Liu \MakeLowercase{\textit{et al.}}: Multidimensional Graph Neural Networks for Wireless Communications}

\IEEEpubid{0000--0000/00\$00.00~\copyright~2023 IEEE}

\maketitle

\begin{abstract}
Graph neural networks (GNNs) can improve the efficiency of learning wireless policies by leveraging their permutation properties and topology prior. While mismatched permutation property to a policy may degrade the learning performance and overlooked permutations incurs low sample efficiency, there is still lacking a systematical approach for modeling graph and designing structure of GNNs to harness all permutation properties. Moreover, the information of input feature may lose during updating hidden representations with GNNs, which leads to poor learning performance. In this paper, we propose a unified framework to learn permutable wireless policies with multidimensional GNNs, which update the hidden representations of hyper-edges to avoid the information loss. We provide a method to construct graph for a policy, over which a GNN with proper parameter sharing can exploit all possible permutations of the policy. We also investigate the permutability of wireless channels that affects the sample efficiency, and show how to trade off the training, inference, and design complexities of GNNs. To showcase how to design the GNNs within the framework, we consider precoding optimization in different systems. Simulation results validate the gain of the proposed GNNs over existing counterparts from exploiting the permutation prior and avoiding the information loss.
\end{abstract}


\section{Introduction}
Resource allocation and signal processing policies play key roles in supporting high spectral efficiency (SE) and energy efficiency (EE) of wireless systems, which are often designed by solving optimization problems. For instance, to optimize the hybrid of analog and baseband precoding in millimeter wave (mmWave) multiple-input multiple-output (MIMO) systems, various numerical algorithms have been proposed, say manifold optimization (MO), phase extraction method (PEM) and orthogonal matching pursuit (OMP) \cite{MO,OMP}, which are hard to be implemented in real-time due to the high computational complexity.

Learning-based technique is overtaking numerical algorithms, say in achieving good performance with low inference time and reducing the signaling overhead for acquiring accurate channel information. Again taking hybrid precoding as an example, a fully-connected neural network (FNN) was used to learn both analog and baseband precoders for a single user (SU)-MIMO system  in \cite{DNN2019}. Convolutional neural networks (CNNs) were designed to learn the analog precoder for multi-user (MU)-MIMO systems in \cite{HybridPrecodingCNN}, to learn both analog and baseband precoders in \cite{BF2021}, and to learn hybrid precoding in wideband SU-MIMO system with implicit channel estimation in \cite{learnOFDM}. Extensive results  have demonstrated the benefits of the learning-based solutions.

Graph neural networks (GNNs) have recently been introduced to learn diverse wireless policies, say power control/allocation \cite{Eisen,shenyifei,GJ,heterD2D}, user/link scheduling \cite{Embedding_Lee2021,RISscheduling,GBLinks, IOT}, access point (AP) selection/user association \cite{APselection,heterAssoc}, and precoding  \cite{RISGNN,ZBC,Korea}. This is motivated by their advantages of achieving better performance with fewer training samples than FNNs and CNNs, learning over graphs with different sizes, and decentralized inference \cite{review2021,LYsurvey,jiang2022graph}, which originate in leveraging prior knowledge. In addition to topology prior, GNNs can also exploit another prior: permutation properties of wireless policies.
A GNN with the same permutation property as a policy can learn the policy efficiently, and is possible to be scalable to large-scale problems and generalizable to unseen system scales \cite{Eisen,shenyifei,GJ}. Otherwise, a GNN either cannot well learn the policy \cite{GJ} or requires high sample and space complexities \cite{ZBC}.

\IEEEpubidadjcol
Nonetheless, designing GNNs to perform well with high learning efficiency is challenging, which consists of modeling graphs and designing structures (i.e., update equations). Constructing appropriate graphical models is the premise of applying GNNs.
%the performance and efficiency of GNNs highly depend on the way to model the graph, select the structure, and tune the hyper-parameters. 
%One source of the performance degradation lies in the unmatched permutation properties of a GNN and a wireless policy.
%%that the function space of searching does not contain the policy.
By updating hidden representation over graphs, the topology information can be harnessed naturally by GNNs. Yet how to exploit the permutation prior with GNNs is far from well-understood.  Existing works only consider a specific policy with special permutation equivariance (PE) property \cite{GJ,ZBC,RISscheduling,RISGNN} or several policies with the same PE property \cite{shenyifei, Eisen}. Due to the lack of generic approach of identifying permutation properties of a policy and modeling  graph, some permutation prior is often overlooked, leading to large  hypothesis space and hence high sample complexity. Moreover, previous works only consider PE and permutation invariance properties, whereas many wireless policies have more complex permutation properties, which are far beyond the one-dimensional (1D)-PE \cite{RISGNN}, two-dimensional (2D)-PE \cite{ZBC}, joint-PE \cite{shenyifei, Eisen}, or their combinations \cite{GJ}.

Except \cite{ZBC}, all existing studies learn wireless policies with vertex-GNNs \cite{Eisen,shenyifei,GJ,heterD2D,RISscheduling,Embedding_Lee2021,IOT,GBLinks,APselection,heterAssoc,RISGNN}, where the hidden representations of vertices are updated by aggregation and combination in each layer. Vertex-GNNs were designed in these works because ``vertex-level'' tasks were considered, i.e., the tasks where the output variables
of the problems are defined on vertices (e.g., power control in \cite{Eisen,shenyifei}). In  \cite{ZBC}, an edge-GNN was designed because learning to precode is an  ``edge-level'' task where the output variables are defined on edges.

It has been found in \cite{howPowerful} that vertex-GNNs may suffer from weak expressive power, i.e., the GNNs perform worse for classification tasks due to unable to distinguish some graphs.
When learning wireless policies, the weak expressive power comes from \emph{information loss}: the useful information of the input of a policy is compressed by a GNN during updating hidden representation.
%for a policy are not identical for different s, but the policy learned by a GNN yields the same output for these inputs.
Take precoding as an example, where the precoding matrices for different channel matrices are not identical.
However, different channel coefficients will become indistinguishable due to the dimension compression after aggregation  (consisting of processing and pooling) at vertices if a vertex-GNN with linear processing is used to learn the precoding policy over a graph with antenna and user vertices.
To improve the expressive power of a vertex-GNN, aggregation function should be injective \cite{howPowerful}. This can be realized by using FNNs with sufficiently wide output layer for processing, as designed in \cite{shenyifei,Korea} but without explanations. Yet such vertex-GNNs are hard to be trained, and it is unknown how wide the output layer of the FNN-processors should be.
%In \cite{RISGNN}, the precoding is learned by a vertex-GNN with vertex-update, where only users are modeled as vertices each with channel vector as feature. Since the precoding vector of each user can be obtained by the updated channel vector, the channel information does not lose but only partial prior knowledge is leveraged.
The information loss can be avoided by updating the hidden representations of edges when learning some policies, say the precoding policy in \cite{ZBC},
%with which the information of each individual edge can be preserved during combination. %because the combining function for an edge can preserve the individual information of the edge.
%However,
but simply using edge-GNNs cannot avoid the information loss for all policies, say hybrid precoding. In fact, the information loss has never been
mentioned in the literature of intelligent communications.
%, where the channel coefficient between each antenna and each antenna is updated.

In this paper, we strive to design efficient GNNs for learning wireless policies by avoiding information loss and exploiting permutation prior. Since the information loss comes from the dimension compression in the update procedure, we propose a multidimensional (MD)-GNN framework that updates the hidden representations of hyper-edges in the space spanned by all the input and output tensors of a problem. The framework consists of input layer, update layers, and output layer, which respectively play the role of increasing the input dimension, updating the hidden representations of hyper-edges, and decreasing the dimension of hidden representation for yielding output. To reveal the potential in reducing sample complexity by GNNs, we show how to find all permutation properties of a policy. Noticing the fact that the permutability of a policy comes from the underlying sets in the optimization problem to obtain the policy and the statistics of input variables of the policy, we show how to identify the sets in optimization problems and analyze the permutability of input variables. Because the permutability of the functions representable by a GNN depends on the types of vertices and the structure of the GNN, we provide an approach to construct a graph, which relates the sets of a problem to the vertices in the graph such that the GNN with judiciously designed update equation of each layer is with matched permutation property to the policy. To accommodate diverse wireless policies, we classify
the problems into two categories and take representative examples to elaborate on how to  model graphs and design MD-GNNs.
%Basically, all permutations should be considered from the indexes of features in a problem, which are also different sets of entities and decide different types of vertices on a graph. Thus, each feature can be located on vertices, edges, or hyper-edges on the graph, by their indexes. Then, updating hidden representation of tensors with maximal dimension is necessary to avoid information loss.
The main contributions are listed as follows.
\begin{itemize}
\item[$\bullet$] We propose a generic MD-GNN framework for wireless policies aimed to avoid information loss and exploit permutation prior. To achieve the first goal, we update the hidden representation tensors of hyper-edges. To achieve the second goal, we provide a generic approach to identify permutation properties of a policy and construct graphs from optimization problems, and design the MD-GNN to satisfy the proved permutable conditions. While we use examples such as hybrid precoding to introduce the framework for easy exposition, the MD-GNNs are applicable to both edge-level and vertex-level tasks or their hybrid.

\item[$\bullet$] We address two practical issues that have never been discussed in literature. One is the permutability of channel samples, which affects the permutability of precoding policies. Another is the tradeoff among the training,  inference, and design complexities of GNNs, which can be achieved by intentionally giving up some permutations.
%To the best of our knowledge, these issues .
%    They can achieve comparable performances with the commonly used iterative algorithm, which are more preferable than FNN and CNN. Meanwhile, they require much fewer training samples and free weights than FNN and CNN to obtain the same performance.
\end{itemize}

Different from previous works where vertex-GNNs were designed with heuristically constructed graphs for some policies with several special PE properties \cite{Eisen,shenyifei,GJ} or without considering PE property \cite{Embedding_Lee2021,IOT,GBLinks}, we exploit all permutation properties of a variety of wireless policies and model graphs systematically. Different from \cite{shenyifei,Korea,heterD2D,GBLinks,IOT} that improve learning performance empirically by increasing the width of output layer of FNN-processor, we avoid dimension compression by updating hyper-edge representation. Different from the edge-GNNs designed for edge-level tasks in literature (e.g., \cite{ZBC}),
we update edge/hyper-edge representations for avoiding information loss by pre-determining the dimensions of different layers.

The rest of the paper is organized as follows. Section \ref{sec:graph} introduces several notions and classifies the permutable problems. Sections \ref{sec:indep} and \ref{sec:dep}  propose the MD-GNN framework for the permutable problems with independent and dependent permutations.
%Section \ref{sec:indep} proposes the MD-GNN framework for the permutable problems with independent permutations.
%Section \ref{sec:dep} extends the framework to the permutable problems with dependent permutations.
Section \ref{practical} addresses the two practical issues. Section \ref{sec:result} provides simulations, and Section \ref{sec:conclu} concludes the paper.

\emph{Notations:}
%Lower case letters denote scalars. Lower case bold-faced letters  denote column vectors. Upper case bold-faced letters denote matrices and tensors.
$\!\left\|\cdot\right\|_2$ denotes the $\ell_2$-norm of a vector. $(\cdot)^T$, $(\cdot)^H\!$, and $\left\|\cdot\right\|_F$ denote transpose, Hermitian transpose, and Frobenius norm of a matrix, respectively. ${\bf I}$ denotes identity matrix. $({\bf X})_{i,j,k}$ denotes an element with index $i,j,k$ in an order-three tensor $\bf X$, and ${\rm vec}({\bf X})=$ $[({\bf X})_{1,1,1}, ({\bf X})_{1,1,2}, \cdots,$ $ ({\bf X})_{1,1,N_1}, ({\bf X})_{1,2,1}, \cdots, ({\bf X})_{N_1,N_2,N_3}]^T \in \mathbb{C}^{N_1N_2N_3}$ stands for the vectorized  tensor ${\bf X} \in \mathbb{C}^{N_1\times N_2\times N_3}$, which are also applicable to tensors with other orders.

\section{Sets, Graphs, Permutable Problems and Policies}\label{sec:graph}

\begin{figure*}[!htb]
	\centering
	\includegraphics[width=0.8\linewidth]{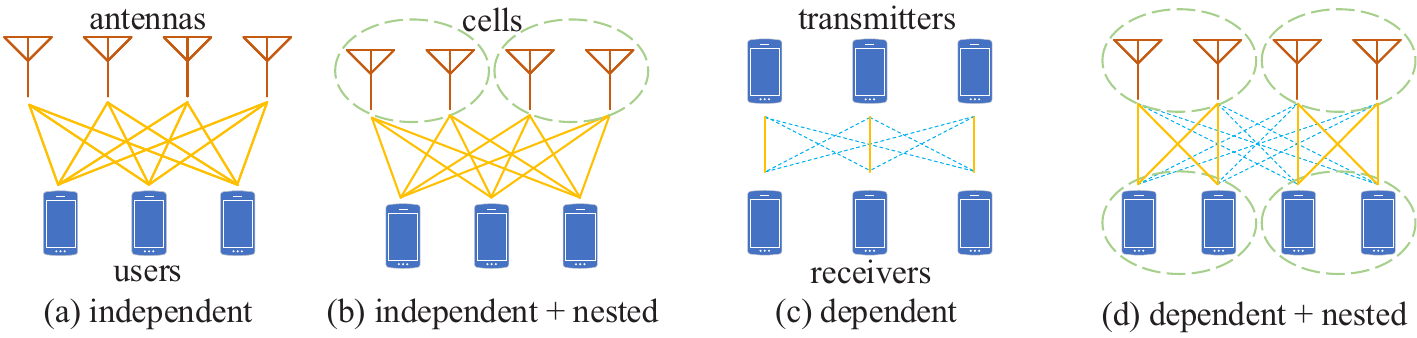}
	\vspace{-2mm}
	\caption{Illustration of the two categories of permutable problems. (a) Precoding in a MISO system\cite{ZBC}, where a BS with four antennas serves three users.
		(b) Precoding in a CoMP-JT system \cite{compjt}, where two BSs each with two antennas jointly serve three users.
		(c) Power control in interference channel \cite{shenyifei}, which consists of three transceiver pairs.
		(d) Power allocation in a cellular system \cite{GJ}, where each of the two BSs only serves the associated two users. }
	\label{fig:fig-example}	\vspace{-4mm}
\end{figure*}

In this section, we first introduce the notions to be used throughout the paper, including sets, graphs, feature tensors,  wireless policies, permutable problems, permutable functions, and permutable policies. Then, we classify the permutable problems into two categories.

\emph{Set and permutation:} A set consists of unordered elements. A nested set is a set of sub-sets. The elements in a set or a sub-set, and the sub-sets in a nested set can be permuted arbitrarily.
The elements in different sets or different nested sets can be permuted independently or dependently (i.e., permuted jointly), depending on the relation between the sets or the nested sets.
%based on whether the permutation of elements in one set has influence on those in another set. The permutation of elements in a set can be called as ``the permutation of a set'' for short.

\emph{Graph and feature tensor:} A graph is composed of vertices, edges, and the associated features. An edge is usually connected with two vertices. If an edge is connected with more than two vertices, then it is a hyper-edge.  A vertex or edge may be associated with feature. The features of all vertices and all edges
(or all hyper-edges) in a graph are called vertex-feature and edge-feature (or hyper-edge-feature)
for short, respectively, which can be expressed as tensors. Representing features by tensors can reserve the high dimensional form of the data.
An order-$M$ tensor ${\bf X} \in \mathbb{C}^{N_1\times N_2 \cdots \times N_M}$ or ${\bf X} \in \mathbb{R}^{N_1\times N_2 \cdots \times N_M}$  has $M$ dimensions (and $M$ indices).
%, which is also called the $M$th-order tensor with dimensions $(N_1, N_2, \cdots N_M)$.
For example, an order-two tensor (i.e., a matrix) has two dimensions, where the rows and the columns are respectively its first and second dimensions.
%Vectors are order-one tensors, and matrices are order-two tensors.

A graph may consist of more than one type of vertices. The vertices of the same type constitute a set or a nested set.
%A Vertex or edge may have feature.
%The features of all vertices in one type are referred to as a vertex-feature, recorded as a vector usually. The features of all edges (hyper-edges) connecting same two (more than two) types of vertices are referred to as a edge-feature (hyper-edge-feature), recorded as a matrix (tensor) usually.
The indices of the vertices correspond to the indices of the elements in a feature vector, the rows or columns  in a feature matrix, or the dimensions in a feature tensor, which are re-ordered accordingly with the permutation of the vertices.
%If the orders of vertices are permuted, all elements in associated vectors will move to new places, we said the vectors are permuted or acted by permutations. Denote $\overline{{\bf H}}_1$ as the result of ${\bf H}_1$ permuted by associated permutations, and so do other features.
When vertex-feature or edge-feature is expressed as a vector or matrix, a permutation matrix ${\bf \Pi}$ can be used to represent the permutation. When a feature is expressed as higher-order tensor, using $\pi(\cdot)$, which maps the $i$-th element in a set into the $\pi(i)$-th element in the set, to represent the permutation of the elements in each dimension of the tensor is more concise.  For example, for the edges connecting $N_1$ vertices of the first type and $N_2$ vertices of the second type, if their features can be expressed as a matrix ${\bf E}$, permuting the matrix into $\overline{{\bf E}}={\bf \Pi}^T_1{\bf E}{\bf \Pi}_2$ can be expressed as $(\overline{{\bf E}})_{i,j} = ({\bf E})_{\pi_1(i),\pi_2(j)},i=1,\cdots,N_1, j=1,\cdots,N_2$, where $[\pi_k(1), \cdots, \pi_k(N_k)]^T ={\bf \Pi}_k^T[1,\cdots,N_k]^T, k=1,2$.

\emph{Wireless policy:} A resource allocation or signal processing policy in wireless communications can usually be expressed as a multivariate function $({\bf W}_1,\cdots, {\bf W}_O)=f({\bf E}_1,\cdots, {\bf E}_I)$, where ${\bf W}_1, \cdots, {\bf W}_O$ represent optimization variables, ${\bf E}_1, \cdots, {\bf E}_I$ represent  known environment parameters.
The policy can be obtained from the following constrained optimization problem
\vspace{-1mm}
	\begin{align*}
	P0: \!\! \max_{{\bf W}_1, \cdots\!, {\bf W}_O} & g_0({\bf W}_1, \cdots\!, {\bf W}_O, {\bf E}_1, \cdots\!, {\bf E}_I) \notag\\
	{\rm s.t.} & g_i({\bf W}_1, \!\cdots\!,  {\bf W}_O, {\bf E}_1, \!\cdots\!, {\bf E}_I)\geq 0, i=1,\!\cdots\!,N_c,
	\end{align*}
where $g_0(\cdot)$ is the objective function, and $g_i(\cdot)$ is the $i$-th constraint function.
%, and $N_c$ is the number of constraints.

Both ${\bf W}_i,i=1,\cdots,O$ and ${\bf E}_i,i=1,\cdots,I$ can be expressed as tensors, called output tensors and input tensors, respectively. Denote $\overline{{\bf W}}_i$ and $\overline{{\bf E}}_i$ as the permuted versions of ${\bf W}_i$ and ${\bf E}_i$  with \emph{all possible permutations} induced by the sets in a problem, as to be explained later.

\emph{Permutable function:} It is the function defined on sets, whose input-output mapping remains unchanged after  its input and output tensors are permuted with \emph{all possible permutations}. For example, a function defined on one set ${\bf y}=f({\bf x})$ has 1D-PE property (i.e., ${\bf \Pi}_1^T{\bf y}=f({\bf \Pi}_1^T{\bf x})$), a function defined on two sets ${\bf Y}=f({\bf X})$ has 2D-PE property (i.e., ${\bf \Pi}_1^T{\bf Y}{\bf \Pi}_2 =f( {\bf \Pi}_1^T{\bf X}{\bf \Pi}_2)$) or joint-PE property (i.e., ${\bf \Pi}_1^T{\bf Y}{\bf \Pi}_1 =f( {\bf \Pi}_1^T{\bf X}{\bf \Pi}_1)$), where $\bm \Pi_1$ and $\bm \Pi_2$ are permutation matrices.

\emph{Permutable problem:} It is the optimization problem whose  objective function and constraints are permutable functions. For example, $P0$ is a permutable problem when $g_i(\overline{{\bf W}}_1, \cdots, \overline{{\bf W}}_O, \overline{{\bf E}}_1, \cdots, \overline{{\bf E}}_I) = g_i({\bf W}_1, \cdots, {\bf W}_O, {\bf E}_1, \cdots, {\bf E}_I), i=0,1,\cdots,N_c$.

\emph{Permutable policy:} A feasible policy of a permutable problem is a permutable policy if it is a permutable function.
If $P0$ is a permutable problem, then the resulting permutable policies will satisfy the following permutation property: $(\overline{{\bf W}}_1,\cdots, \overline{{\bf W}}_O)=f(\overline{{\bf E}}_1,\cdots, \overline{{\bf E}}_I)$.

More than one optimal policy may be obtained from a permutable problem \cite{LSJ}. If only one optimal policy can be obtained from a permutable problem, then it is a permutable policy \cite{Eisen}. Otherwise, it is not hard to show that  at least one optimal policy is permutable.

Permutation property of a policy is a kind of prior knowledge, which can be exploited to improve the learning efficiency of GNNs by enforcing them to learn permutation functions after the GNNs are
judiciously designed. To provide a unified framework for learning a variety of wireless policies with GNNs, we divide permutable problems into two categories.

In the first category, all sets in a problem can be permuted independently, as illustrated in Fig. \ref{fig:fig-example}(a)(b).
In Fig. \ref{fig:fig-example}(a), there are two sets in the precoding policy for the multi-input-single-output (MISO) system: antenna-set and user-set. Permuting antennas and users independently does not change the policy.
In Fig. \ref{fig:fig-example}(b), the four antennas constitute a nested set including two sub-sets, each consisting of the antennas in each base station (BS) of the coordinated multi-point with joint transmission (CoMP-JT) system. The three users constitute a set. Permuting antennas and users independently does not change the policy.

In the second category, some of the sets must be permuted dependently, as illustrated in Fig. \ref{fig:fig-example}(c)(d). In Fig. \ref{fig:fig-example}(c), the policy will remain unchanged only if the transmitters and receivers are permuted jointly, i.e., their permutations are dependent. In Fig. \ref{fig:fig-example}(d), both antennas and users constitute nested sets, each including two sub-sets. Each sub-set of antennas consists of the two antennas in each BS, and each sub-set of users consists of the two users in each cell.
The policy will remain unchanged only if the antennas and users  in the same cell are permuted jointly.

\section{A GNN Framework: All Sets Permuted Independently}\label{sec:indep}
In this section, we propose a MD-GNN to avoid information loss and exploit permutation prior when learning policies from the problems in the first category. We first introduce a systematical method of identifying sets and constructing a graph from a permutable problem. Then, we consider precoding in downlink MISO systems, as a kind of representative problems illustrated in Fig. \ref{fig:fig-example}(a). Finally, we consider other problems in the category illustrated in Fig. \ref{fig:fig-example}(b).

\subsection{Method for Identifying Sets and Modeling Graphs}\label{modelgraph}
%To harness all possible permutations of a permutable policy, we model a graph for a GNN to learn the policy as follows.
To model a graph such that a GNN can exploit permutation prior of a permutable policy, we find all sets in a problem and then define the vertices according to the sets.

The permutability of a problem and permutation properties of the resulting policy are induced by sets, over which the objective and constraint functions of the problem are defined. To identify all sets in a problem, we can find all the dimensions of the input and output tensors of the problem, and regard the elements in each dimension as a set. If the objective and constraints are unchanged after permuting these elements, which is true for commonly considered (e.g., SE- or EE-maximal) problems, then they indeed constitute a set.

The permutability of the functions representable by a GNN is induced by the vertices of a graph over which the GNN learns. To design a GNN with matched permutation properties to a policy, we relate the number of sets to the number of vertex types, and define the elements in each set of a problem as the vertices of each type. Then, by observing the dimensions of every input and output tensors, we can identify all the edges and all the features of vertices and edges.
%, as to be exemplified later.

\subsection{MD-GNN for Optimizing Precoding in MU-MISO Systems}
\subsubsection{Hybrid Precoding Policy and its Permutation Property}
As a motivating example for designing a MD-GNN, we consider baseband and analog precoding in mmWave MU-MISO system, where a BS with $N_t$ antennas  and $N_s$ RF chains serves $K$ single-antenna users. The two precoders can be jointly optimized, say from the following weighted sum-rate maximization problem that takes into account of user fairness \cite{HybridPrecodingCNN},\vspace{-1mm}
\begin{subequations}
	\begin{align}
	{P1}: &\max_{{\bf W}_{RF},{\bf W}_{BB}} \notag\\
	\sum_{k=1}^K & \beta_k \log_2 \left(1\!+\!\frac{|{\bf h}^H_k {\bf W}_{RF} {\bf w}_{BB_k}|^2}{\sum_{i=1,i\neq k}^K |{\bf h}^H_k {\bf W}_{RF} {\bf w}_{BB_i}|^2+\sigma^2} \!\right)\label{objective1}\\
	{\rm s.t.} ~~
	&  ||{\bf W}_{RF} {\bf W}_{BB} ||^2_F = P_{tot},\label{eq:constraint2}\\
	& |\left({\bf W}_{RF}\right)_{j,l}|=1,j=1,\cdots,N_t,l=1,\cdots,N_s\label{eq:constraint3},
	\end{align}
\end{subequations}
where ${\bf W}_{RF} \in \mathbb{C}^{N_t\times N_s}$ is the analog precoder, ${\bf W}_{BB}=[{\bf w}_{BB_1},\cdots,{\bf w}_{BB_K}]\in \mathbb{C}^{N_s \times K}$ is the baseband precoder, $P_{tot}$ is the total power, $\beta_k, k=1,\cdots, K$ are the weights controlling the fairness among users, ${\bf h}_k \in \mathbb{C}^{N_t}$ is the channel vector of the $k$-th user, and $\sigma^2$ is the noise power. \eqref{eq:constraint2} is the power constraint, and \eqref{eq:constraint3} is the constant modulus constraint for the analog precoder.

Denote  $({\bf W}_{RF}, {\bf W}_{BB})$ as a feasible solution of problem $P1$, which satisfies the constraints but may not achieve the maximal weighted sum-rate.
Denote a feasible precoding policy as $({\bf W}_{RF}, {\bf W}_{BB}) = f({\bf H},{\bm \beta},P_{tot})$,
where $f: \mathbb{C}^{K \times N_t} \times \mathbb{R}^K \times \mathbb{R} \rightarrow \mathbb{C}^{N_t \times N_s} \times \mathbb{C}^{N_s \times K}$ is a mapping, ${\bf H}=[{\bf h}_1,\cdots,{\bf h}_K]^T \in \mathbb{C}^{K\times N_t}$, and ${\bm \beta}=[\beta_1,\cdots,\beta_K]^T \in \mathbb{R}^K$. In order for the learned hybrid precoding policy being adaptive to different channels, user fairness criteria, and BSs with different maximal powers, the environment parameters of the policy include ${\bf H}$, ${\bm \beta}$, and $P_{tot}$.
%The policy consists of two kinds of optimization variables and three kinds of environmental parameters (i.e., $O=2, I=3$).

%By definition, a function can be a one-to-one mapping or many-to-one mapping.

As proved in \cite{LSJ}, $P1$ is a permutable problem of  user-set, antenna-set, and RF chain-set. The elements in each set can be permuted
arbitrarily and the three sets can be permuted independently.
The three  sets are identified by observing all the dimensions of the input and output matrices of the mapping $f(\cdot)$ and examining if
%permuting the users, antennas and RF chains independently will change the objective function and constraints. A
a feasible solution $({\bf W}_{RF},{\bf W}_{BB})$ of $P1$  for ${\bf H},{\bm \beta},P_{tot}$ achieves the same weighted sum-rate with $(\overline{{\bf W}}_{RF},\overline{{\bf W}}_{BB})$ for $\overline{{\bf H}}, \overline{{\bm \beta}},P_{tot}$   \cite{LSJ}, where
\begin{eqnarray}\label{3set-PE}
&&\overline{{\bf W}}_{RF} \triangleq {\bf \Pi}_2^T{\bf W}_{RF}{\bf \Pi}_{3}, ~~
\overline{{\bf W}}_{BB}  \triangleq {\bf \Pi}_{3}^T{\bf W}_{BB}{\bf \Pi}_1, ~~ \notag\\
&&\overline{{\bf H}}  \triangleq {\bf \Pi}_1^T{\bf H}{\bf \Pi}_2,~~ \overline{{\bm \beta}}  \triangleq {\bf \Pi}_1^T{\bm \beta},
\end{eqnarray}
are respectively the permuted version of ${\bf W}_{RF},{\bf W}_{BB}, {\bf H},{\bm \beta}$ with \emph{all three possible permutations}, ${\bf \Pi}_1$, ${\bf \Pi}_2$, and ${\bf \Pi}_3$  respectively represent the  permutations on users, antennas, and RF chains.
%A permutable policy can be obtained from the problem.

The existence of ${\bf \Pi}_3$ indicates that $N_s!$ equivalent feasible solutions can be obtained for one group of environment parameters ${\bf H},{\bm \beta},P_{tot}$ by permuting the $N_s$ RF chains. The \emph{equivalency} is in the sense of achieving the same weighted sum-rate.
This indicates that there exist $N_s!$ feasible hybrid precoding policies that are equivalent, since a one-to-many mapping is not a function by definition.
When ${\bf\Pi}_3={\bf I}$, one feasible policy satisfies $\left( {\bf\Pi}_2^T {\bf W}_{RF}, {\bf W}_{BB}{\bf\Pi}_1 \right)=f\left({\bf \Pi}_1^T{\bf H}{\bf \Pi}_2,{\bf \Pi}_1^T{\bm \beta},P_{tot} \right)$.
%\begin{equation}\label{eq:taskfucPE1}
%\left( {\bf\Pi}_2^T {\bf W}_{RF}, {\bf W}_{BB}{\bf\Pi}_1 \right)=f\left({\bf \Pi}_1^T{\bf H}{\bf \Pi}_2,{\bf \Pi}_1^T{\bm \beta},P_{tot} \right).
%\end{equation}
The existence of ${\bf\Pi}_1$ and ${\bf\Pi}_2$ means that ${K!N_t!}$ equivalent feasible solutions can be obtained from the policy for ${K!N_t!}$  channel matrices permuted from one channel matrix.
When ${\bf\Pi}_3\neq {\bf I}$, other feasible policies satisfy
\begin{equation}\label{eq:taskfucPE2}
\!\! \left( {\bf\Pi}_2^T {\bf W}_{RF}{\bf\Pi}_3, {\bf\Pi}_3^T {\bf W}_{BB}{\bf\Pi}_1 \! \right) \!\!=\!\! f_{{\bf\Pi}_3} \!\! \left({\bf \Pi}_1^T{\bf H}{\bf \Pi}_2,{\bf \Pi}_1^T{\bm \beta},P_{tot} \! \right)\!\! , \!\!\!
\end{equation}
where $f_{{\bf\Pi}_3}(\cdot)$ denotes $N_s!-1$ permutable functions.  As shown in \eqref{eq:taskfucPE2},
${\bf \Pi}_3$ is not associated with any environment parameters, but introduces a relation between ${\bf W}_{RF}$ and ${\bf W}_{BB}$.

%It is shown from \eqref{3set-PE},  \eqref{eq:taskfucPE1} and \eqref{eq:taskfucPE2} that there are
In summary, $N_s!$ permutable hybrid precoding policies can be obtained from problem $P1$, each satisfies a \emph{three-set permutation property}: $\left( \overline{\bf W}_{RF}, \overline{\bf W}_{BB} \right)=f\left(\overline{\bf H},\overline{\bm \beta},P_{tot} \right)$.
%This is not a kind of PE property \cite{GJ}.
%, since the input matrix or vector and each output matrix are not with the same permutation.
%By taking this prior into consideration, a deep learning algorithm only needs to find a function satisfying the property, which is more efficient.

%Without taking the permutational properties in \eqref{eq:taskfucPE1} and \eqref{eq:taskfucPE2} into consideration, a deep learning algorithm has to find a large number of equivalent feasible precoding policies each yielding a large number of equivalent feasible solutions, which is inefficient.

%\begin{proposition}
%$(\overline{{\bf W}}_{RF},\overline{{\bf W}}_{BB})$ is a feasible solution to environment parameter $(\overline{{\bf H}}, \overline{{\bf \sigma}},\overline{{\bm \beta}},P_T)$, which achieves the same weighted sum-rate as $({\bf W}_{RF},{\bf W}_{BB})$ to environment parameter $({\bf H}, {\bf \sigma},{\bm \beta},P_T)$.
%\end{proposition}
%\begin{proof}
%See Appendix A.
%\end{proof}

%The problem decides what sets it has, namely how many kinds of vertices it have, but we must find them out first, substitute all features and verify that the problem is permutable second. It is possible to forget some sets. This is why most of existing works only take users into account.

\subsubsection{3D-GNN for Learning the Hybrid Precoding Policy}\label{subsec:3dgnn}
%To learn the permutable hybrid precoding policy in the MU-MISO system efficiently, a GNN can be used.
%, where the input is $({\bf H},{\bm \sigma},{\bm \beta},P_T)$, and the output is $({\bf W}_{RF},{\bf W}_{BB})$.
\begin{figure*}[!htb]
	\centering
	\subfigure[Narrow-band MU-MISO]{
		\begin{minipage}[c]{0.33\linewidth}
			\centering
			\includegraphics[width=1\linewidth]{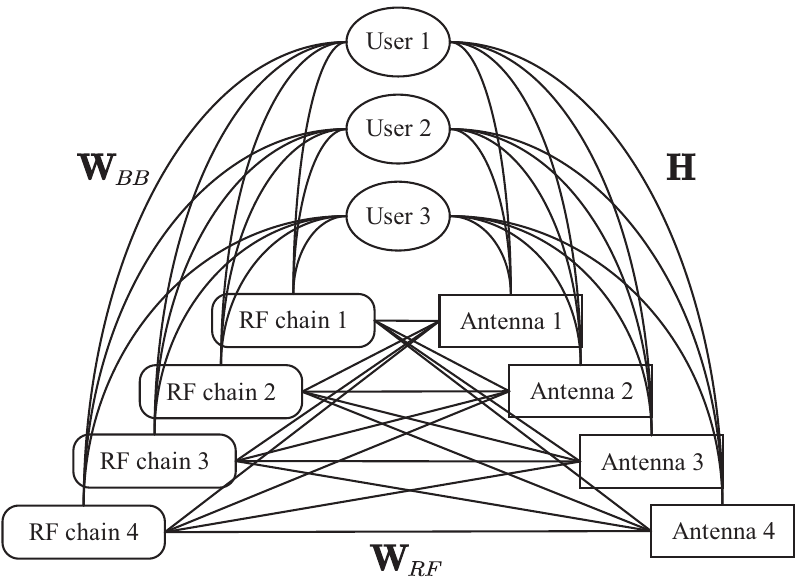}
		\end{minipage}
	}
	\hspace{-25pt}
	\subfigure[Wideband MU-MISO]{
		\begin{minipage}[c]{0.33\linewidth}
			\centering
			\includegraphics[width=1\linewidth]{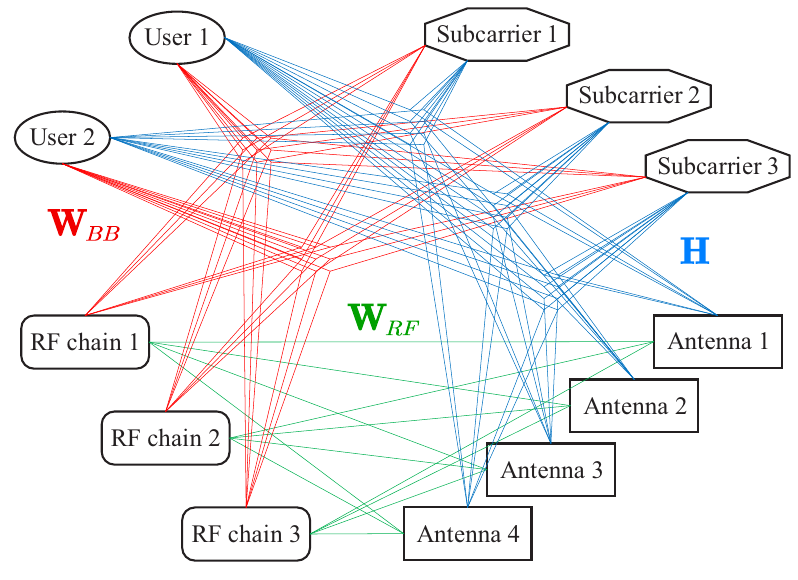}
		\end{minipage}
	}
	\hspace{-15pt}
	\subfigure[Narrow-band MU-MIMO]{
		\begin{minipage}[c]{0.33\linewidth}
			\centering
			\includegraphics[width=1\linewidth]{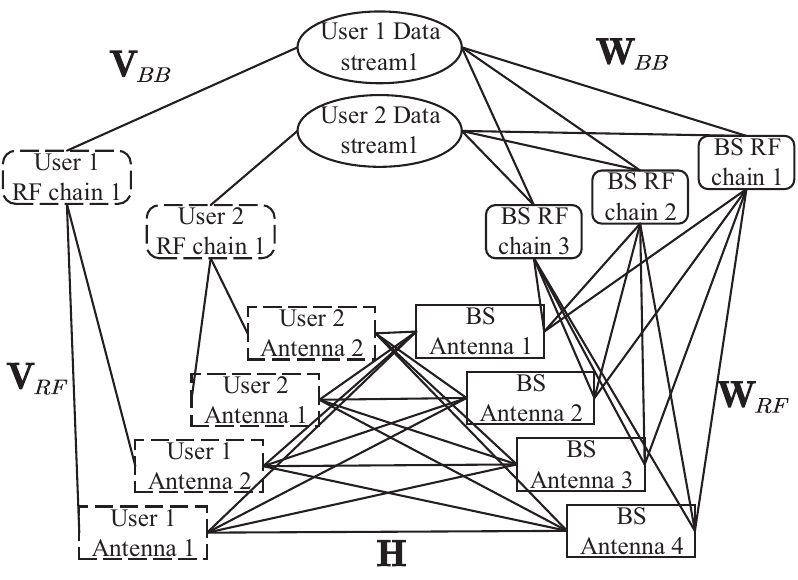}
		\end{minipage}
	}
	\vspace{-2mm}
	\caption{Graphs for learning analog and baseband precoding. In (a) and (b), all permutations are independent. In (c), the permutations of data stream, RF chains at users, and the antennas at users are partially dependent.}
	\label{fig:fig-connect}\vspace{-4mm}
\end{figure*}

To learn a permutable hybrid precoding policy by a GNN with the three-set permutation property, we construct a graph with three types of vertices: user-vertices, antenna-vertices (AN-vertices for short), and RF chain-vertices (RF-vertices for short), as illustrated in Fig. \ref{fig:fig-connect}(a), which come from three sets.

%There is an edge between two vertices of different types, say between each user-vertex and each antenna-vertex.
From the dimensions of ${{\bf W}}_{RF}, {{\bf W}}_{BB}, {\bf H}, {\bm \beta}$, and $P_{tot}$, we can identify the edges of the graph and the features of the vertices and edges. For example, ${\bf H}\in \mathbb{C}^{K\times N_t}$ is with user dimension and antenna dimension, and hence consists of the features of all the edges connecting user-vertices and AN-vertices. Similarly, we can see that ${\bf W}_{RF} \in \mathbb{C}^{N_t\times N_s}$ consists of the features of all the edges between AN-vertices and RF-vertices, and ${{\bf W}}_{BB}\in \mathbb{C}^{N_s\times K}$ consists of the features of all the edges between RF-vertices and user-vertices. ${\bm \beta} \in \mathbb{R}^K$ is only with user dimension, and hence consists of the features of user-vertices. $P_{tot}$ is not a feature of any vertex or any edge. Finally, the AN- and RF-vertices are without features.

%The permutations on the features can be expressed as $\overline{{\bm \beta}} = {\bf \Pi}_1^T{\bm \beta}$, $\overline{{\bf H}} = {\bf \Pi}_1^T{\bf H}{\bf \Pi}_2$, $\overline{{\bf W}}_{BB} = {\bf \Pi}_{3}^T{\bf W}_{BB}{\bf \Pi}_1$, and $\overline{{\bf W}}_{RF} = {\bf \Pi}_2^T{\bf W}_{RF}{\bf \Pi}_{3}$, where ${\bf \Pi}_1$, ${\bf \Pi}_2$ and ${\bf \Pi}_3$ are the pairwise permutations on user-set, antenna-set, and RF chain-set, respectively.

When learning over the constructed graph, the input feature ${\bf H}\in \mathbb{C}^{K\times N_t}$ is a matrix with user and antenna dimensions, ${\bm \beta}\in \mathbb{R}^K$ is a vector with user dimension, and $P_{tot}$ is a scalar. All input features span a 2D space with dimensions of ${K\times N_t}$. However, ${\bf W}_{RF} \in \mathbb{C}^{N_t\times N_s}$ is a matrix with antenna and RF chain dimensions and ${\bf W}_{BB} \in \mathbb{C}^{N_s \times K}$ is with RF chain and user dimensions, i.e., all output features span a three-dimensional (3D) space with dimensions of ${K\times N_t \times N_s}$.
%that are the projections respectively on two of the three dimensions of an order-three tensor in ${K\times N_t \times N_s}$ complex space.
%In fact, the root cause of the existence of one-to-many mapping shown in \eqref{eq:taskfucPE2} is that,
%the input space is with less dimensions than the output space.
Since ${{\bf W}}_{RF}$ and ${{\bf W}}_{BB}$ are features on edges, it seems natural to design an edge-GNN. Yet if we use the edge-GNN in \cite{ZBC} to learn the policy over the graph, i.e., the representations of ${{\bf W}}_{RF}$ and ${{\bf W}}_{BB}$ are updated alternatively in each hidden layer, then
the channel matrix will still be compressed after aggregation. This is
 because ${\bf H}$, ${{\bf W}}_{RF}$, and ${{\bf W}}_{BB}$
% the information of user dimension and antenna dimension of ${\bf H}$ are respectively compressed when updating the hidden representation for ${{\bf W}}_{RF}$ and ${{\bf W}}_{BB}$.
lie in different 2D spaces, which are respectively the projections on different dimensions in the 3D feature space. %, whereas the aggregated results are on the intersection lines of every two 2D spaces.
%, unless the number of pooling functions is equal to $\max \{N_t, K\}$ \cite{CCB2020}.
To avoid losing the information of input features, we propose a framework of 3D-GNN  for the three-set problem, where the hyper-edge representations are updated in the 3D feature space spanned by all the input and output features. The framework consists of an input layer to increase the dimension of input feature space, update layers to learn the hyper-edge representations, and an output layer to project the representation in the last update layer into ${{\bf W}}_{RF}$ and ${{\bf W}}_{BB}$, i.e.,
\begin{subequations}\begin{align}
&{\rm input~layer:~~} {\bf X}_1 = f_{in}({\bf H},{\bm \beta},P_{tot}; {\bf a}), \label{update eqa}\\
&{\rm update~layers:~~} {\bf X}_{l+1} = f_l({\bf X}_l ), l = 1,\cdots,L-1, \label{update eqb}\\
&{\rm output~layer:~~} ({\bf W}_{RF}, {\bf W}_{BB})= f_{out}({\bf X}_L),\label{update eqc}
\end{align} \end{subequations}
where ${\bf a}=[a_1,\cdots,a_{N_s}]^T\in\mathbb{R}^{N_s}$ is a virtual feature vector introduced for increasing the dimension of input features as explained soon, and $L$ is the number of layers of the GNN. Each element of ${\bf X}_1$ is the feature of a hyper-edge connecting one user-, one AN-, and one RF-vertices. ${\bf X}_l\in \mathbb{R}^{C_l\times K\times N_t \times N_s}$ is the hidden representation in the $l$-th layer, and $C_l$ is the number of ``channels'' in the $l$-th layer. The notion of ``channel'' is the same as that of CNNs, while the ``channels''  are not permutable  with vertices in GNN. Since the real and imaginary parts of ${\bf H}$ and the three input features ${\bf H},{\bm \beta},P_{tot}$ are not permutable  with vertices, $C_1 = 2+1+1 = 4$. Since the real and imaginary parts of ${\bf W}_{BB}$ and ${\bf W}_{RF}$ are also not permutable, $C_L=4$. Both $L$ and $C_l, l=2,\cdots,L-1$ are hyper-parameters.
We refer to ${\bf X}_l$ as \emph{a 3D-feature despite that it is  an order-four tensor}, because the first dimension is irrelevant to permutations.

Denote $\overline{{\bf X}}_l$ as the permuted version of the tensor ${\bf X}_l$ with three independent permutations ${\bf \Pi}_1$, ${\bf \Pi}_2$, and ${\bf \Pi}_3$,  where $(\overline{{\bf X}}_l)_{c,k,n_t,n_s} = ({\bf X}_l)_{c,\pi_1(k),\pi_2(n_t),\pi_3(n_s)}$.
The following proposition provides the condition for the 3D-GNN to satisfy the permutation property of the precoding policy.
%Associating $ {\bf X}_l$ with all three permutations is necessary to ensure that the permutations of inputs can control those of the outputs.
%Since ${\bf X}_l, l=1,\cdots, L$ contain all three permutations

\begin{proposition}\label{prop:composite}
{\bf (Permutable condition)}: If $f_{in}(\cdot)$, $f_{out}(\cdot)$, and $f_l(\cdot), l = 1,\cdots,L-1$ are permutable functions that respectively satisfy the following permutation properties,
\begin{subequations}
	\begin{align}
	&  \overline{\bf X}_1 = f_{in}(\overline{\bf H},\overline{\bm \beta},P_{tot}; \overline{\bf a}),\label{6a}\\
	&  \overline{\bf X}_{l+1} = f_l(\overline{\bf X}_l ), l = 1,\cdots,L-1,\label{6b}\\
	& (\overline{\bf W}_{RF}, \overline{\bf W}_{BB})= f_{out}(\overline{\bf X}_L),\label{6c}
	\end{align}
\end{subequations}
then the policy learned by the 3D-GNN, $({\bf W}_{RF}, {\bf W}_{BB})= f_{out}(f_{L-1}\cdots f_1(f_{in}({\bf H},{\bm \beta},P_{tot};{\bf a})))$, is a permutable function, which satisfies the three-set permutation property.
\end{proposition}
\begin{proof}
See Appendix \ref{app:A}.
\end{proof}

In what follows, we show how to design the layers that satisfy the properties in \eqref{6a}-\eqref{6c}.

\begin{figure*}[htb]
	\centering
	\includegraphics[width=0.85\linewidth]{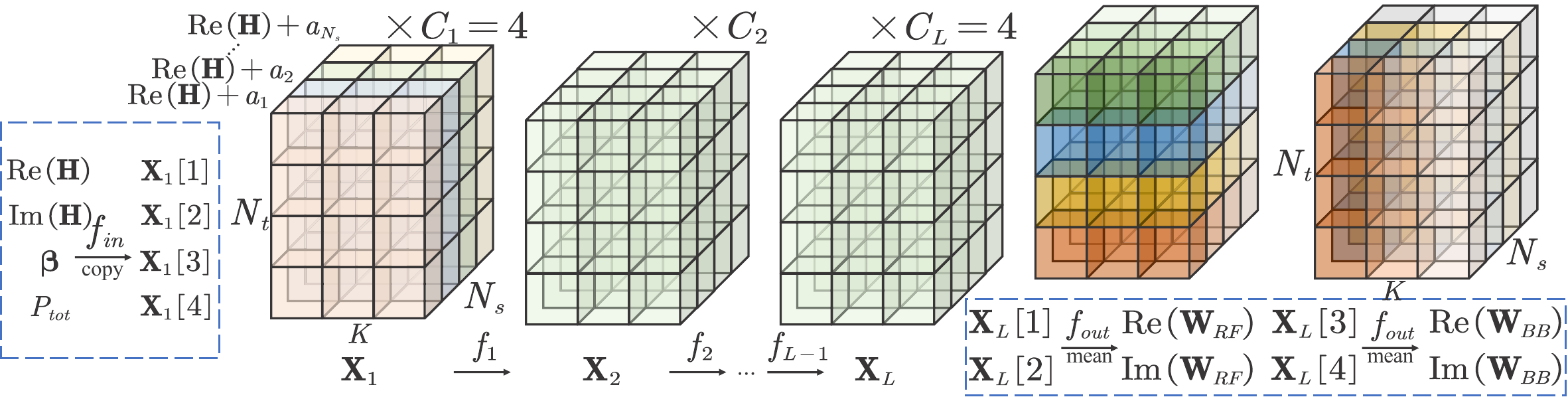}
	\vspace{-2mm}
	\caption{The input, update, and output layers of the 3D-GNN for hybrid precoding.} \label{fig:fig-3dGNN}
	\vspace{-4mm}
\end{figure*}

{\bf Input Layer:} The role of  $f_{in}(\cdot)$ is to map the environment parameters into hyper-edge-feature in the 3D feature space, where ${\bf X}_1\in \mathbb{R}^{C_1\times K\times N_t \times N_s}$. To this end, we can arrange $\rm{Re}({\bf H})$, $\rm{Im}({\bf H})$, ${\bm \beta}$, and $P_{tot}$ respectively in the first, second, third, and fourth ``channels'' (denoted as ${\bf X}_1[1]$, ${\bf X}_1[2]$, ${\bf X}_1[3]$, and ${\bf X}_1[4]$).
Then, we represent the input feature in each ``channel'' as an order-three tensor. This can be accomplished by copying each kind of feature along the dimensions that it does not have. In particular, since the dimension of RF chains is missing in ${\bf H}$, we copy the real part and imaginary part of matrix ${\bf H}$ along the RF chain dimension for $N_s$ times. Similarly, we copy the user-vertex feature ${\bm \beta}$ along the antenna and RF chain dimensions, and copy the scalar $P_{tot}$ along the user, antenna, and RF chain dimensions.
To increase the dimension of input feature space meanwhile do not occupy an extra ``channel'', we add the virtual-feature vector ${\bf a}$
on the replicas of at least one input feature, say $\rm{Re}({\bf H})$ as shown in Fig. \ref{fig:fig-3dGNN}. The vector ${\bf a}$ can be generated randomly before
training and is the same for every training sample and test sample.
%The randomly generated values of $a_i$ should be comparable with ${\rm Re}({\bf H})$.
Without introducing ${\bf a}$, the 3D-GNN will output identical results over the RF chain dimension in every hidden representation ${\bf X}_l$ and in the outputs (i.e., ${\bf X}_l$ and the output features lie in the 2D space with dimensions $K\times N_t$), which incurs unacceptable learning performance.

 $f_{in}(\cdot)$ consists of the ``arranging'' operation and the ``copying'' operation with the added virtual vector. It can be verified that $f_{in}(\cdot)$ is a permutable function satisfying the property in \eqref{6a}.

{\bf Output Layer:}
$f_{out}(\cdot)$ maps ${\bf X}_L$ into $({\bf W}_{RF}, {\bf W}_{BB})$, which can simply be accomplished by averaging ${\bf X}_L$ over different dimensions.
In particular, we take the average of the first and second ``channels'' of ${\bf X}_L$ over user dimension to obtain the real and imaginary parts of ${\bf W}'_{RF}$, and take the average of the third and fourth ``channels'' over antenna dimension to obtain the real and imaginary parts of ${\bf W}'_{BB}$.
%, i.e.,
%\begin{equation}\label{outWW}
%{\bf W}'_{RF} = {\mathop{\rm mean}_{K}} \left({\bf X}_L[1:2]\right),{\bf W}'_{BB} = {\mathop{\rm mean}_{N_t}} \left({\bf X}_L[3:4]\right),
%\end{equation}
%where ${\bf X}_L[a:b]$ means the operation that extracts the elements from $a$ to $b$ in the first dimension of ${\bf X}_L$.
To satisfy the constraints in problem $P1$, we project ${\bf W}_{RF}^\prime$ and ${\bf W}_{BB}^\prime$ into
$({\bf W}_{RF})_{j,l} = \left. {({{\bf W}_{RF}^\prime})_{j,l}}\middle/ {|({{\bf W}_{RF}^\prime})_{j,l}|} \right., j=1,\cdots,N_t,l=1,\cdots,N_s$ and ${\bf W}_{BB} = \left.{\sqrt{P_{tot}}{{\bf W}_{BB}^\prime}}\middle/\|{\bf W}_{RF} {\bf W}_{BB}^\prime\|_F\right.$.

$f_{out}(\cdot)$ consists of the ``averaging'' and the ``projection'' operations. It can be verified that $f_{out}(\cdot)$ is a permutable function  satisfying the property in \eqref{6c}.

It is worthy to note that the proposed framework is applicable to the problems with other constraints (e.g., the quality of service constraint in \cite{antennaselection}), by using existing methods to deal with complex constraints (e.g., Lagrange dual learning method \cite{SCJ}).

{\bf Update layers:} The design of $f_l(\cdot)$ is flexible. One approach is first to satisfy the 3D-PE property in \eqref{6b} by introducing parameter sharing into a FNN using existing methods (e.g.,  \cite{IandE,acrosssets}) and then harnessing topology prior by setting some weights as zero. Another approach is to design a proper GNN that updates representations of hyper-edges in the 3D space, where both topology and permutation priors can be leveraged  implicitly by the constructed graph.
%We can also design attention-based update equations, as exemplified later.
%can be simply linear functions cascaded with activation functions. This corresponds to using a linear function for aggregating (i.e., linear processor and sum pooling) and . These functions can be replaced by nonlinear functions like FNNs.

To help understand how the GNN structure is designed for embedding the two kinds of priors, we consider the first approach. In order to satisfy the property in \eqref{6b} in each update layer by using the methods in \cite{IandE,acrosssets}, the update equation can be expressed as ${\rm vec}({{\bf X}_{l+1}}) = \sigma ({\bf P}_l{\rm vec}({{\bf X}_l})),l=1,\cdots,L-1$, where ${\rm vec}({{\bf X}_l}) \in \mathbb{R}^{KN_tN_s}$ is the vectorized version of tensor ${\bf X}_l\in \mathbb{R}^{ K\times N_t \times N_s}$, and ${\bf P}_l$ is the structured weight matrix in the $l$-th layer to be designed. For notational simplicity, we ignore the first dimension of hidden representation consisting of multiple ``channels'' that are irrelevant to the permutations,
 when we discuss update layers.
After omitting the element-wise activation function that does not affect permutation properties, the update equation becomes ${\rm vec}({{\bf X}_{l+1}}) = {\bf P}_l{\rm vec}({{\bf X}_l})$. To satisfy \eqref{6b}, i.e., ${\rm vec}(\overline{\bf X}_{l+1}) = {\bf P}_l{\rm vec}(\overline{\bf X}_l)$ (which can be re-written as $({\bf \Pi}_1^T\otimes {\bf \Pi}_2^T\otimes {\bf \Pi}_3^T){\rm vec}({\bf X}_{l+1}) ={\bf P}_l ({\bf \Pi}_1^T\otimes {\bf \Pi}_2^T\otimes {\bf \Pi}_3^T){\rm vec}({\bf X}_l)$), a fixed-point equation $({\bf \Pi}_1^T\otimes {\bf \Pi}_2^T\otimes {\bf \Pi}_3^T){\bf P}_l={\bf P}_l({\bf \Pi}_1^T\otimes {\bf \Pi}_2^T\otimes {\bf \Pi}_3^T)$ can be obtained by replacing ${\rm vec}({\bf X}_{l+1})$ with ${\bf P}_l{\rm vec}({{\bf X}_l})$. The weight matrices have been found from the equation with the method in \cite{IandE}, which have the following structure with three-level hierarchical parameter sharing\cite{acrosssets}, 
\begin{align}\label{PS-P}
&{\bf P}_l = \begin{pmatrix}
{\bf P}_{l,1} & {\bf P}_{l,2} & \cdots & {\bf P}_{l,2} \\
{\bf P}_{l,2} & {\bf P}_{l,1} & \cdots & {\bf P}_{l,2} \\
\vdots & \vdots & \ddots & \vdots \\
{\bf P}_{l,2} & {\bf P}_{l,2} & \cdots & {\bf P}_{l,1}
\end{pmatrix} \in \mathbb{R} ^{KN_tN_s\times KN_tN_s} , \notag\\
&{\bf P}_{l,m} = \begin{pmatrix}
{\bf P}_{l,m,1} & {\bf P}_{l,m,2} & \cdots & {\bf P}_{l,m,2} \\
{\bf P}_{l,m,2} & {\bf P}_{l,m,1} & \cdots & {\bf P}_{l,m,2} \\
\vdots & \vdots & \ddots & \vdots \\
{\bf P}_{l,m,2} & {\bf P}_{l,m,2} & \cdots & {\bf P}_{l,m,1}
\end{pmatrix} \in \mathbb{R} ^{N_tN_s\times N_tN_s},\nonumber \\
&{\bf P}_{l,m,r} = \begin{pmatrix}
p_{l,m,r,1} & p_{l,m,r,2} & \cdots & p_{l,m,r,2} \\
p_{l,m,r,2} & p_{l,m,r,1} & \cdots & p_{l,m,r,2} \\
\vdots & \vdots & \ddots & \vdots \\
p_{l,m,r,2} & p_{l,m,r,2} & \cdots & p_{l,m,r,1}
\end{pmatrix} \in \mathbb{R} ^{N_s\times N_s},\notag\\
& m=1,2,r=1,2.
\end{align}

\begin{figure*}[htb]
	\centering
	\includegraphics[width=0.7\linewidth]{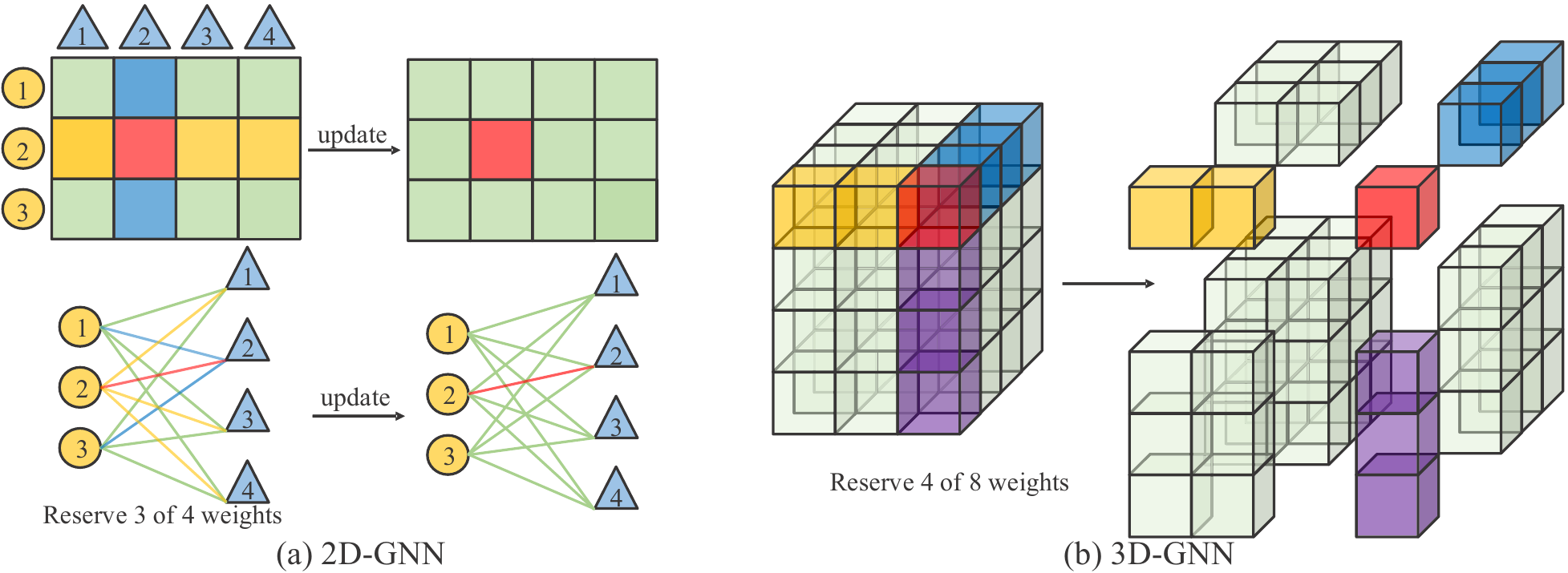}
	\vspace{-2mm}
	\caption{An update layer of 2D- and 3D-GNNs, where a square represents an element in ${\bf X}_l$. In (a), the red element is updated by summing itself with the blue elements and the yellow elements in ${\bf X}_{l-1}$ using different weights. We only reserve the three of four weights (not sum the green elements), because the green edges are not adjacent to the red edge in the corresponding bipartite graph below. In (b), the red element is updated by summing itself with other three kinds of elements  using different weights. The three kinds of elements only have one index difference from the red element.}\label{fig:fig-update}
	\vspace{-4mm}
\end{figure*}

To further harness topology prior, only the information in adjacent hyper-edges of the hyper-edge whose representation is updated should be aggregated at each layer, where the adjacent hyper-edges have two common vertices with the hyper-edge. Hence, the weights for non-adjacent hyper-edges should be zero. % i.e., $p_{l,m,r,s}=0$ if there are two or more ``2'' in the indices $m,r,s$.
After exploiting the 3D-PE property, there are $2^3$ different weights (i.e. trainable parameters) in ${\bf P}_l$. After further exploiting the topology information, there are only $3+1=4$ weights in the weight matrix of each layer of the GNN,
%which are the weights for the updated hyper-edge  itself and the hyper-edges with only one different index,
as illustrated in Fig. \ref{fig:fig-update}. For a general $M$-set problem, only $M+1$ of $2^M$ weights need to be trained for each ``channel''.

To help understand what are adjacent and non-adjacent edges for updating the  hidden representation of an edge, we also provide an update layer for 2D-GNN in Fig. \ref{fig:fig-update}(a) whose hidden representations of edges can be expressed as a matrix.
 %(i.e., order-two tensor) and the weight matrices only have two hierarchies of parameter sharing.}
 Since there are $C_l$ ``channels'' in the $l$-th layer, there are $4C_lC_{l+1}$ weights  between the $l$-th and $(l+1)$-th layers.
%, where $L$ and $C_l, L=2,\cdots, L-1$ are hyper-parameters.
%Similarly, the $D+1$ free weights are those whose indexes have at most one ``2''.
 %are trained by back propagation algorithms, where $L$ and $C_l, l=2,\cdots,L-1$ can be arbitrarily chosen. The larger $L$ and $C_l$, the stronger representation capability of GNNs, similarly to CNNs.

{\bf Remark 1:} The precoding problem in \cite{ZBC} can be regarded as a special case of problem $P1$ with $N_s=N_t$ and ${\bf W}_{RF}={\bf I}$.  If we only consider channel as environment parameter, then the precoding policy is ${\bf W}_{BB}^T=f({\bf H})$, which  satisfies ${\bf \Pi}_1^T{\bf W}_{BB}^T{\bf \Pi}_2=f({\bf \Pi}_1^T{\bf H}{\bf \Pi}_2)$. A 2D-GNN can be used to learn this policy over a graph only consisting of user-vertex and AN-vertex, which updates edge representations ${\bf X}_l \in \mathbb{R}^{C_l\times K\times N_t}$ (i.e., the edge-GNN in \cite{ZBC}). The weight matrices in the update layers have the same structure as in \eqref{PS-P} but with two-level hierarchical parameter sharing. After harnessing  topology prior, the representation of the edge between the $k$-th user and the $n$-th antenna is updated by ${\bf x}_{l+1,k,n} = \sigma({\bf P}_{l,1,1}{\bf x}_{l,k,n}+{\bf P}_{l,2,1}\sum_{i=1,i\neq k}^{K}{\bf x}_{l,i,n}+{\bf P}_{l,1,2}\sum_{m=1,m\neq n}^{N_t}{\bf x}_{l,k,m})$.
We can also design an attention-based 2D-GNN, whose update equation is ${\bf x}_{l+1,k,n} = \sigma({\bf P}_{l,1,1}{\bf x}_{l,k,n} +\sum_{i=1,i\neq k}^{K}{\bm \alpha}_{l,k,i}\odot {\bf P}_{l,2,1}{\bf x}_{l,i,n} +  {\bf P}_{l,1,2}\sum_{m=1,m\neq n}^{N_t}{\bf x}_{l,k,m}/N_t)$, where ${\bm \alpha}_{l,k,i} = \sum_{n=1}^{N_t} {\bf Q}_l {\bf x}_{l,k,n}  \odot{\bf R}_l {\bf x}_{l,i,n}/N_t$, ${\bf Q}_l$ and ${\bf R}_l$ are trainable weights, and $\odot$ denotes Hadamard product. The 2D-GNN with this update equation is referred to as A2D-GNN, which differs from GAT that is a vertex-GNN \cite{GAT}. We can prove that A2D-GNN satisfies the 2D-PE property. Such an attention mechanism can also be introduced to other MD-GNNs.

{\bf Remark 2:} We can show that the problem of jointly optimizing the precoding of an $N_t$-antenna AP and the $M$ reflection coefficients of a reconfigurable intelligent surface (RIS) in a MU-MISO system in \cite{RISGNN} is permutable, which is with user-set, AN-set, and RIS element-set. There exists a permutable policy, denoted as  $({\bf W}, {\bf \Theta})=f({\bf H}, {\bf G},{\bf B})$, where ${\bf W}\in \mathbb{C}^{N_t\times K}$ is the precoding matrix, ${\bf \Theta} \in  \mathbb{C}^{M\times M}$ is a diagonal matrix consisting of all the reflection coefficients, ${\bf H} \in \mathbb{C}^{K\times N_t}$, ${\bf G} \in \mathbb{C}^{K\times M}$, and ${\bf B} \in \mathbb{C}^{M\times N_t}$ are respectively the channel matrices from the AP to users, from the RIS to users, and from the AP to the RIS,
and $K$ is the number of users. The policy satisfies another \emph{three-set property}: $({\bf \Pi}_2^T{\bf W}{\bf \Pi}_1, {\bf \Pi}_3^T{\bf \Theta}{\bf \Pi}_3)=f({\bf \Pi}_1^T{\bf H}{\bf \Pi}_2,{\bf \Pi}_1^T{\bf G}{\bf \Pi}_3, {\bf \Pi}_3^T{\bf B}{\bf \Pi}_2)$.
By using the method in section \ref{modelgraph}, we can establish a graph with three types of vertices: users, antennas, and RIS elements, for learning this policy. ${\bf \Theta}$ consists of the features of the RIS-vertices. ${\bf H}$ and ${\bf W}$ are the features of the edges between user- and AN-vertices, ${\bf G}$ is the feature of the edges between user- and RIS-vertices, and ${\bf B}$ is the feature of the edges between RIS- and AN-vertices. The output features ${\bf \Theta}$ and ${\bf W}$ are defined on vertices and edges, respectively. Hence, this is a hybrid of vertex-level and edge-level task.
We can use a 3D-GNN to learn this policy, with hidden representation ${\bf X}_l \in \mathbb{R}^{C_l \times K\times N_t \times M}$. In the input layer, $C_1=6$, and we respectively copy the real and imaginary parts of $\bf H$, $\bf G$, and $\bf B$ along RIS element dimension, antenna dimension, and user dimension. Since the input features span a 3D space with dimensions of ${K\times N_t \times M}$, which is the same as the space spanned by the output features, the virtual feature ${\bf a}$ is no longer required. In the output layer, $C_L=4$, $\bf W$ and $\bf \Theta$ are respectively obtained by first averaging ${\bf X}_L$ over RIS element dimension and by averaging ${\bf X}_L$ over user dimension and antenna dimension, and then by projection to satisfy the constraints. ${\bf P}_l, l=1,\cdots, L-1$ in the update layers are with the same structure as in \eqref{PS-P}.

\subsubsection{4D-GNN for Learning Wideband Hybrid Precoding Policy}\label{subsubsec:OFDM}
Consider a mmWave MU-MISO-orthogonal frequency division multiplexing (OFDM) system with $M$ subcarriers. Then, the SE-maximal wideband hybrid precoding problem (refers to as $P2$ in the sequel) only differs from $P1$ in the baseband precoder, where ${\bf W}_{BB}=[{\bf W}_{BB}^1,\cdots,{\bf W}_{BB}^M] \in \mathbb{C}^{M \times N_s\times K}$,
%\begin{subequations}
%	\begin{align*}
%	P2: &\max_{{\bf W}_{RF},{\bf W}_{BB}} ~~ \frac{1}{M}\sum_{m=1}^M\sum_{k=1}^K \log_2 \left(1+ \gamma_k^m\right), \\
%	{\rm s.t.} ~~
%	&  \sum_{m=1}^M ||{\bf W}_{RF} {\bf W}_{BB}^m ||^2_F = P_{tot},~~ |\left({\bf W}_{RF}\right)_{j,l}|=1,j=1,\cdots,N_t,l=1,\cdots,N_s,
%	\end{align*}
%\end{subequations}
%where ${\bf W}_{RF} \in \mathbb{C}^{N_t\times N_s}$ is the analog precoder, ${\bf W}_{BB}=[{\bf W}_{BB}^1,\cdots,{\bf W}_{BB}^M] \in \mathbb{C}^{M \times N_s\times K}$ is the baseband precoder,
and ${\bf W}_{BB}^m=[{\bf w}_{BB_1}^m,\cdots,{\bf w}_{BB_K}^m]\in \mathbb{C}^{N_s\times K}$  is the baseband precoder on the $m$-th subcarrier.
%and $\gamma_k^m= \frac{|{{\bf h}^m_k}^H {\bf W}_{RF} {\bf w}_{BB_k}^m |^2}{{\sum_{i=1,i\neq k}^K |{{\bf h}^m_k}^H {\bf W}_{RF} {\bf w}_{BB_i}^m|} ^2+\sigma^2}$.

Denote a feasible precoding policy as $({\bf W}_{RF}, {\bf W}_{BB}) = f({\bf H})$,
where ${\bf H}=[{\bf H}^1,\cdots,{\bf H}^M] \in \mathbb{C}^{M \times K\times N_t}$, ${\bf H}^m=[{\bf h}_1^m,\cdots,{\bf h}_K^m]^T \in \mathbb{C}^{K\times N_t}$ consists of the channel vectors of all users on the $m$-th subcarrier, and $f: \mathbb{C}^{M \times K \times N_t} \rightarrow \mathbb{C}^{N_t \times N_s} \times \mathbb{C}^{M \times N_s \times K}$ is a mapping. Again, we only take ${\bf H}$ as environment parameter for notational simplicity. From the dimensions in the mapping, we can see that this problem is with  user-set, antenna-set, RF chain-set, and subcarrier-set. When the  four sets are respectively permuted independently by $\pi_1(\cdot)$, $\pi_2(\cdot)$, $\pi_3(\cdot)$, and $\pi_4(\cdot)$, it is easy to show that a feasible solution of ${\bf W}_{BB}$ and ${\bf W}_{RF}$ for ${\bf H}$ achieves the same SE with $\overline{\bf W}_{BB}$ and $\overline{\bf W}_{RF}$ for $\overline{\bf H}$, where
%. the permuted version of ${\bf H}$, ${\bf W}_{BB}$ and ${\bf W}_{RF}$  can be represented by
$(\overline{\bf W}_{BB})_{m,n_s,k}=({\bf W}_{BB})_{\pi_4(m),\pi_3(n_s),\pi_1(k)}$, $\overline{\bf W}_{RF}={\bf \Pi}_2^T{\bf W}_{RF}{\bf \Pi}_3$, and $(\overline{\bf H})_{m,k,n_t} = ({\bf H})_{\pi_4(m),\pi_1(k),\pi_2(n_t)}$.  In other words, there exists a permutable precoding policy that satisfies a \emph{four-set permutation property}: $(\overline{\bf W}_{RF}, \overline{\bf W}_{BB}) = f(\overline{\bf H})$.
%, and $\overline{\bf W}_{BB}$ and $\overline{\bf H}$ are defined in \eqref{3set-PE}.

To harness four possible permutations, we establish a graph with user-vertices, AN-vertices, RF-vertices, and subcarrier-vertices, as illustrated in Fig. \ref{fig:fig-connect}(b), all without features. The precoding matrix ${\bf W}_{RF}$ consists of the features on the edges between  RF-vertices and AN-vertices. The tensors ${\bf W}_{BB}$ and ${\bf H}$ are the features on the hyper-edges connecting RF-, subcarrier-, and user-vertices and  the hyper-edges connecting AN-, subcarrier-, and user-vertices, respectively.

We can use a 4D-GNN to learn the policy, where ${\bf X}_l \in \mathbb{R}^{C_l\times M \times K\times N_t\times N_s}$, ${\bf H}$  is copied along the RF chain dimension and its real part is added with a virtual feature in the input layer. Before harnessing topology information, the weight matrices in update layers have similar structure to those in \eqref{PS-P} but with four-level hierarchical parameter sharing. In the output layer, ${\bf W}_{BB}$ and ${\bf W}_{RF}$ are obtained by first averaging ${\bf X}_L$ over antenna dimension and averaging ${\bf X}_L$ over user and subcarrier dimensions, respectively, and then by projection to satisfy the constraints.

{\bf Remark 3:}  A wideband baseband precoding policy ${\bf W}_{BB}=f({\bf H})$ can be obtained from $P2$ by setting $N_s=N_t$ and ${\bf W}_{RF}={\bf I}$, where ${\bf H}, {\bf W}_{BB} \in \mathbb{R}^{2\times M\times K\times N_t}$. This policy satisfies a \emph{three-set property}: 3D-PE, which can be learned by a 3D-GNN with ${\bf X}_l \in \mathbb{R}^{C_l\times M\times K\times N_t}$ over a graph consisting of user-vertices, AN-vertices and subcarrier-vertices.  This 3D-GNN is without dimension increase or dimension decrease in the input and output layers, which differs from the 3D-GNN for learning the hybrid precoding policy from $P1$. 

\subsection{Application of the MD-GNN to Other Problems with Independent Sets}
In the sequel, we provide two problems in the first category illustrated in Fig. \ref{fig:fig-example}(a) other than precoding.
Consider a SU-MIMO system, where a BS with $N_t$ transmit (TX)-antennas serves a single user with $N_r$ receive (RX)-antennas, and $M$ pilots are used for channel estimation.
\subsubsection{MIMO Signal Detection}
The received signal is ${\bf y} = {\bf H}{\bf s}+{\bf n}$, where ${\bf y} \in \mathbb{C}^{N_r}$, ${\bf H} \in \mathbb{C}^{N_r \times N_t}$ is the channel matrix, ${\bf s}\in \mathbb{C}^{N_t}$ is the transmitted signal, and ${\bf n} \in \mathbb{C}^{N_r}$ is the noise. The signal can be detected by solving a problem, say $\min_{{\bf s}}  ||{\bf y}- {\bf H}{\bf s}||_2^2$ \cite{detection}.
%\begin{subequations}
%	\begin{align*}
%	P4: \min_{{\bf x}} ~~&||{\bf y}- {\bf H}{\bf x}||_2^2,~~
%	{\rm s.t.} ~~  {\bf x} \in \mathcal{X}.
%	\end{align*}
%\end{subequations}

It is not hard to show that this is a permutable problem with RX-antenna-set and TX-antenna-set, where the corresponding permutations are ${\bf \Pi}_1$ and ${\bf \Pi}_2$, respectively. The detection policy, denoted as ${\bf s}=f({\bf H}, {\bf y})$, satisfies a \emph{two-set property}$: {\bf \Pi}_2^T{\bf s}=f({\bf \Pi}_1^T{\bf H}{\bf \Pi}_2, {\bf \Pi}_1^T{\bf y})$.

We establish a graph with two types of vertices: TX- and RX-antennas, which are respectively with features ${\bf s}$ and ${\bf y}$. $\bf H$ is the feature of the edges between RX- and TX-vertices.
We can use a 2D-GNN to learn this policy, whose weight matrix in each update layer has the same structure as ${\bf P}_l$  in \eqref{PS-P} but with two-level hierarchical parameter sharing before setting the weights of non-adjacent edges as zero. In the input layer, ${\bf X}_1\in \mathbb{R}^{4\times N_r\times N_t}$ is composed of $\bf H$ and $\bf y$ copied along the TX-antenna dimension, and the virtual feature $\bf a$ is unnecessary. In the output layer, $\bf s$ is obtained by averaging ${\bf X}_L\in \mathbb{R}^{2\times N_r\times N_t}$ over RX-antenna dimension followed by a projection function for classification.

\subsubsection{MIMO Channel Estimation}
The received  pilots can be expressed as ${\bf Y}_p= {\bf H}{\bf S}_p+{\bf N}_p$, where ${\bf Y}_p \in \mathbb{C}^{N_r\times M}$ consists of $M$ received pilots, ${\bf S}_p \in \mathbb{C}^{N_t\times M}$  consists of $M$ transmitted pilots, and ${\bf N}_p \in \mathbb{C}^{N_r\times M}$ is the noise. The channel matrix $\bf H$ can be estimated by solving an optimization problem, say $\min_{{\bf H}} ||{\bf Y}_p- {\bf H}{\bf S}_p||_F^2$  \cite{jointdetection}.

This is a permutable problem with RX-antenna-set, whose permutation is $\bf \Pi$. The channel estimation policy ${\bf H}=f({\bf Y}_p)$ satisfies ${\bf \Pi}^T{\bf H}=f({\bf \Pi}^T{\bf Y}_p)$. Both ${\bf H}$ and ${\bf Y}_p$ are vertex-features. A 1D-GNN can learn this policy with hidden representations ${\bf X}_l\in \mathbb{R}^{C_l \times  N_r }$. In the input layer, ${\bf X}_1 = {\bf Y}_p \in \mathbb{R}^{2M\times N_r}$. In the output layer, ${\bf H}={\bf X}_L \in \mathbb{R}^{2N_t\times N_r}$. In each update layer, the weight matrix is with same structure as ${\bf P}_l$  in \eqref{PS-P} but with one-level hierarchical parameter sharing.

\subsection{Extending the MD-GNN for Problems with Independent and Nested Sets}\label{subsec:inde_nest}
Identifying a nested set requires domain knowledge. The weight matrices in update layer of a MD-GNN for the problem with independent and nested sets are no longer with the structure in \eqref{PS-P} since the permutation properties differ. This is shown by the following example problem.

Consider the precoding in a CoMP-JT  system as illustrated in Fig. \ref{fig:fig-example}(b), which consists of $B$ BSs each with $N_t$ antennas jointly serving $K$ users in the $B$ cells \cite{compjt}. The antennas at each BS can be arbitrarily permuted, but the antennas in different BSs cannot, due to the power constraint at each BS. Hence, this problem is with two set: all users constitute a set, while all antennas in the system constitute a nested set and the antennas at each BS constitute a sub-set.

The permutation of the nested set of all antennas can be expressed as ${\bf \Omega}^T\triangleq({\bf \Pi}_3^T\otimes {\bf I}_{N_t}){\rm diag}({\bf \Pi}_{2,1}^T,\cdots,{\bf \Pi}_{2,B}^T)$, where ${\bf \Pi}_{2,b}$ represents the permutations on the antennas in the $b$-th BS and ${\bf \Pi}_3$ represents the permutations on the sub-sets. A precoding policy in CoMP-JT can be obtained from an optimization problem, which is with a set and a nested set. The precoding policy is ${\bf W}_{BB}^T=f({\bf H})$, which satisfies a \emph{nested permutation property}: ${\bf \Pi}^T_1{\bf W}_{BB}^T{\bf \Omega}=f({\bf \Pi}_1^T{\bf H}{\bf \Omega})$, where ${\bf H}\in \mathbb{C}^{ K\times BN_t}$, ${\bf W}_{BB}\in \mathbb{C}^{ BN_t\times K}$, and ${\bf \Pi}_1$ represents the permutations on the users.

The graph for learning this policy can be established as follows: users and antennas are vertices, ${\bf H}$ and ${\bf W}_{BB}$ are features of the edges between the two types of vertices. A 2D-GNN can be used for learning this policy, where ${\bf X}_l\in \mathbb{R}^{C_l\times K\times BN_t}$. Since both ${\bf H}$ and ${\bf W}_{BB}$ are matrices in the same 2D space, it is unnecessary to change the dimensions of the  input and output layers. Hence, an edge-GNN can be used.
In the update layers, the parameter sharing in the update equation ${\rm vec}({{\bf X}_{l+1}}) = \sigma \left({\bf P}_l{\rm vec}({{\bf X}_l})\right)$  can be designed by using the method in \cite{equ} to satisfy the permutation property ${\bf \Pi}^T_1{\bf X}_{l+1}{\bf \Omega}=f_l({\bf \Pi}_1^T{\bf X}_l{\bf \Omega})$.

\section{GNNs: Some Sets Permuted Dependently}\label{sec:dep}
In this section, we extend the MD-GNN for learning the policies from the permutable problems in the second category. In section \ref{subsec:depGNN}, we consider the problems with dependent sets illustrated in Fig. \ref{fig:fig-example}(c). In section \ref{subsec:depnest}, we consider the problems with dependent and nested sets illustrated in Fig. \ref{fig:fig-example}(d). Similar to finding the nested sets, identifying jointly permuted sets using the method in section \ref{modelgraph} also requires domain knowledge. The input, update, and output layers of the extended framework take the same role as the MD-GNN framework for the first category. However, the dimensions of input feature and the last hidden representation are increased and reduced no longer simply by copying and averaging, and the structure of the weight matrix in each update layer needs to be re-designed for the joint permutation property. We still consider the first approach to design the update layer in section \ref{subsec:3dgnn} to embed the two kinds of priors.

%, because some permutations are dependent and then some dimensions are relevant. Walso introduce weight sharing to change their dimensions.

\subsection{Extending the 2D-GNN for the Problems with Dependent Sets}\label{subsec:depGNN}
%\subsubsection{Power Control in Interference Channel}\label{subsec:depGNN}
Consider the power control problem in an interference channel with $K$ single-antenna transceiver pairs in \cite{shenyifei}.
%say $\max_{p_1,\cdots,p_K}  \sum_{k=1}^K \log_2 \left(1+\frac{|h_{k,k}|^2p_k}{\sum_{i=1,i\neq k}^K |h_{k,i}|^2p_i+\sigma^2}\right)~ {\rm s.t.} ~ 0\leq p_k \leq P_M, ~k=1,\cdots,K$\cite{shenyifei},
%%\begin{subequations}
%%	\begin{align*}
%%	P4: \max_{p_1,\cdots,p_K} &\sum_{k=1}^K \log_2 \left(1+\frac{|h_{k,k}|^2p_k}{\sum_{i=1,i\neq k}^K |h_{k,i}|^2p_i+\sigma^2}\right)~
%%	{\rm s.t.} ~ 0\leq p_k \leq P_M, ~k=1,\cdots,K,
%%	\end{align*}
%%\end{subequations}
%where $|h_{k,i}|^2$ is the channel gain between the $k$-th transmitter and the $i$-th receiver, $p_k$ and $P_M$ are the transmit power of the $k$-th transmitter and the maximal power of each transmitter. Denote ${\bf p}=[p_1,\cdots,p_K]^T \in \mathbb{R}^K$, and ${\bf H} \in \mathbb{R}^{K\times K}$ with $({\bf H})_{i,j} = |h_{i,j}|^2$.
This is a two-set permutable problem, where transmitter-set and receiver-set are permuted by a single permutation ${\bf \Pi}$. The policy is ${\bf p}=f({\bf H})$, which satisfies a two-set joint permutation property: ${\bf \Pi}^T{\bf p}=f({\bf \Pi}^T{\bf H}{\bf \Pi})$ (i.e., joint-PE property), where ${\bf p} \in \mathbb{R}^K$ is the transmit power and  ${\bf H} \in \mathbb{R}^{K\times K}$ is the channel gain matrix.

The graph for this problem can be established as follows. Both transmitters and receivers are vertices. The features of the transmitter vertices constitute ${\bf p}$, the receivers have no feature. ${\bf H}$ is the edge-feature between transmitter and receiver vertices.
A 2D-GNN can be used for this vertex-level task, where edge representations
${\bf X}_l \in \mathbb{R}^{C_l\times K\times K}$ are updated. In the input layer, $\bf H$ is arranged in the only one ``channel'' of ${\bf X}_1$ without the need of increasing the dimension.

The update layer ${\rm vec}({{\bf X}_{l+1}}) = \sigma ({\bf P}_l{\rm vec}({{\bf X}_l}))$ should satisfy ${\bf \Pi}^T{\bf X}_{l+1}{\bf \Pi}=f_l({\bf \Pi}^T{\bf X}_l{\bf \Pi})$. To satisfy the joint-PE
property, the structure of ${\bf P}_l$ can be designed by using the method in \cite{IandE}. To further harness the graph topology, the weights in ${\bf P}_l$ for non-adjacent edges are set as zero.

The output layer ${\bf p}=f_{out}({\bf X}_L)$ extracts the information from and reduces the dimension of ${\bf X}_L \in \mathbb{R}^{1\times K\times K}$ to obtain ${\bf p} \in \mathbb{R}^K$.  Since the information in both the second and third dimensions of ${\bf X}_l $ is relevant to ${\bf p} $, we can not reduce the dimension by only taking average over a dimension as in section \ref{sec:indep}. Instead, we design the output layer as ${\bf p}= \sigma ({\bf P}_L{\rm vec}({{\bf X}_L}))$, which satisfies ${\bf \Pi}^T{\bf p}=f_{out}({\bf \Pi}^T{\bf X}_L{\bf \Pi})$. The structure of ${\bf P}_L$ can be designed using the method in \cite{IandE}.

{\bf Remark 4:} The link scheduling policy in D2D networks \cite{Embedding_Lee2021,GBLinks, IOT} and the power control policy in random access systems \cite{Eisen} can also be learned by this 2D-GNN.

\subsection{Extending the MD-GNN for the Problems with Dependent and Nested Sets}\label{subsec:depnest}
\subsubsection{5D-GNN for Learning Hybrid precoding and Combining Policy}
Consider a mmWave MU-MIMO system \cite{HybridPrecodingCNN}, where a BS equipped with $N_t$ antennas and $N_s$ RF chains transmits to $K$ users each receiving $S$ data streams with $N_r$ antennas and $N_v$ RF chains.

For the BS, the analog precoder is ${\bf W}_{RF}\in \mathbb{C}^{N_t\times N_s}$, and the baseband precoder is ${\bf W}_{BB}=[{\bf W}_{BB,1},\cdots,{\bf W}_{BB,K}]\in \mathbb{C}^{N_s\times KS}$. For the $k$-th user, the analog combiner is ${\bf V}_{RF,k}\in \mathbb{C}^{N_v\times N_r}$, and the baseband combiner is ${\bf V}_{BB,k}\in \mathbb{C}^{S\times N_v}$.
Denote ${\bf V}_{RF} = {\rm diag}({\bf V}_{RF,1},\cdots,{\bf V}_{RF,K})\in \mathbb{C}^{KN_v\times KN_r}$, ${\bf V}_{BB} = {\rm diag}({\bf V}_{BB,1},\cdots,{\bf V}_{BB,K})\in \mathbb{C}^{KS\times KN_v}$. ${\bf H} \in \mathbb{C}^{KN_r\times N_t}$ is the channel matrix.

There are five sets in the problem: antennas and RF chains at the BS, data streams, antennas and RF chains at the user. Denote the permutations of BS-antennas and BS-RF chains as ${\bf \Pi}_2$ and ${\bf \Pi}_3$, respectively. The data stream-set, user-antenna-set, and user-RF chain-set are nested sets, and the corresponding permutations are respectively denoted as ${\bf \Omega}_1^T\triangleq({\bf \Pi}_1^T\otimes {\bf I}_S){\rm diag}({\bf \Pi}_{1,1}^T,\cdots,{\bf \Pi}_{1,K}^T)$,  ${\bf \Omega}_2^T\triangleq({\bf \Pi}_1^T\otimes {\bf I}_{N_r}){\rm diag}({\bf \Pi}_{2,1}^T,\cdots,{\bf \Pi}_{2,K}^T)$, and ${\bf \Omega}_3^T\triangleq({\bf \Pi}_1^T\otimes {\bf I}_{N_v}){\rm diag}({\bf \Pi}_{3,1}^T,\cdots,{\bf \Pi}_{3,K}^T)$, all of them partially depend on the permutation of users ${\bf \Pi}_1$, where ${\bf \Pi}_{1,k}$, ${\bf \Pi}_{2,k}$, and ${\bf \Pi}_{3,k}$ are respectively the permutation of data streams, antennas, and RF-chains of the $k$-th user.
Denote the policy as $({\bf W}_{RF},{\bf W}_{BB},{\bf V}_{RF},{\bf V}_{BB})=f({\bf H})$, which satisfies a \emph{five-set joint permutation property}: $({\bf \Pi}_2^T{\bf W}_{RF}{\bf \Pi}_3,$ ${\bf \Pi}_3^T{\bf W}_{BB}{\bf \Omega}_1,$ ${\bf \Omega}_3^T{\bf V}_{RF}{\bf \Omega}_2,$ ${\bf \Omega}_1^T{\bf V}_{BB}{\bf \Omega}_3)=$ $f({\bf \Omega}_2^T{\bf H}{\bf \Pi}_2)$.% Denote $\overline{\bf W}_{RF}={\bf \Pi}_2^T{\bf W}_{RF}{\bf \Pi}_3$, $\overline{\bf W}_{BB}={\bf \Pi}_2^T{\bf W}_{RF}{\bf \Pi}_3$, $\overline{\bf V}_{RF}={\bf \Pi}_2^T{\bf W}_{RF}{\bf \Pi}_3$, $\overline{\bf V}_{BB}={\bf \Pi}_2^T{\bf W}_{RF}{\bf \Pi}_3$, and $\overline{\bf H}={\bf \Omega}_2^T{\bf H}{\bf \Pi}_2$.

The constructed graph is illustrated in Fig. \ref{fig:fig-connect}(c), where BS-antennas, BS-RF chains, data-streams, user-antennas, and user-RF chains are vertices, all without features. $\bf H$ is the edge-feature between BS-antenna and user-antenna vertices, ${\bf W}_{RF}$ is the edge-feature between BS-RF chain  and BS-antenna vertices, ${\bf W}_{BB}$ is the edge-feature between BS-RF chain and data stream vertices, ${\bf V}_{RF}$ is the edge-feature between user-RF chain and user-antenna  vertices, and ${\bf V}_{BB}$ is the edge-feature between user-RF chain and data stream vertices. In this problem, the data stream vertices do not correspond to any physical ``nodes'', whose permutation is  easily  overlooked.

A 5D-GNN can be applied to learn the precoding policy. The input layer is ${\rm vec}({\bf X}_1) = {\bf P}_0 {\rm vec}({\bf H})+{\bf a}$, where  ${\bf X}_1\in \mathbb{R}^{2\times KS \times KN_v \times KN_r \times N_t \times N_s}$, ${\bf a}$ is a virtual feature related to $\bf H$, and the structure of ${\bf P}_0$ should be designed to satisfy $\overline{\bf X}_1=f_{in}({\bf \Omega}_2^T{\bf H}{\bf \Pi}_2)$.
In each update layer, ${\bf X}_l$ is updated by ${\rm vec}({{\bf X}_{l+1}}) = \sigma ({\bf P}_l{\rm vec}({{\bf X}_l}))$ to satisfy the property in \eqref{6b}.
In the output layer, $({\bf W}_{RF},{\bf W}_{BB},{\bf V}_{RF},{\bf V}_{BB})=f_{out}({\bf X}_L)$, which is obtained by first reducing the dimension of ${\bf X}_L$ from eight ``channels'' of ${\bf X}_L$ respectively with four weight matrices (say ${\rm vec}({\bf W}'_{RF})={\bf P}_{L,1}{\rm vec}({\bf X}_L[1,2])$ and ${\rm vec}({\bf W}'_{BB})={\bf P}_{L,2}{\rm vec}({\bf X}_L[3,4])$), and then projecting them to satisfy the constraints. All the weight matrices $ {\bf P}_0$, $ {\bf P}_l$, and $ {\bf P}_{L,i},i=1,2,3,4$ can be designed by first using the method in \cite{equ} and then setting the weights for non-adjacent edges as zero.

\begin{table*}[htb!]
	\centering
	\caption{Wireless policies that can be learned with the MD-GNN Framework}\label{table:problems}
	\vspace{-2mm}
	\footnotesize
	\renewcommand\arraystretch{1}
	\begin{tabular}{l|l|l}
		\hline\hline
		GNN & Independent sets & Dependent sets\\
		\hline
		1D & \makecell[l]{Power allocation among multi-channels \cite{DavidTse} \\ Channel estimation, SU-MIMO \cite{jointdetection}\\ Joint channel estimation and precoding \cite{jointEstimationPrecoding}\\ Bandwidth allocation \cite{bandwidth,SCJ}  } & Not existed \\
		\hline
		2D & \makecell[l]{Precoding, MU-MISO \cite{ZBC, Korea} \\ Antenna selection, MU-MISO \cite{antennaselection} \\ User scheduling, MU-MISO \cite{userscheduling}\\
			Precoding, CoMP-JT \cite{compjt} (nested) \\ Precoding, cell-free MIMO \cite{cellfree} (nested)\\
			Signal detection, SU-MIMO \cite{detection} \\
			Subcarrier assignment \cite{subcarrierAllocation}
			\\Joint channel estimation and signal detection \cite{jointdetection}} & \makecell[l]{Power control, interference channel \cite{shenyifei, SCJ} \\ Link scheduling, D2D \cite{Embedding_Lee2021,GBLinks, IOT}\\
			Power control, random access \cite{Eisen}\\
			Precoding, CoMP-CB \cite{WMMSE} (nested)
		} \\
		\hline
		3D & \makecell[l]{Hybrid precoding, MU-MISO  \cite{HybridPrecodingCNN}\\
			Beamforming, RIS-assisted transmission \cite{RISGNN}\\
			Precoding, SU-MIMO \cite{DavidTse}
		} & \makecell[l]{Partially-connected hybrid precoding, MU-MISO \cite{MO} (nested)\\
			Precoding, MU-MIMO \cite{WMMSE} (nested)
		} \\
		\hline
		4D & \makecell[l]{Hybrid precoding, wideband MU-MISO \cite{yuweiHybridOFDM}
		}& Precoding, wideband MU-MIMO \cite{mu-mimi-ofdm-precode} (nested) \\
		\hline
		5D & Hybrid precoding, SU-MIMO \cite{MO} & Hybrid precoding, MU-MIMO \cite{HybridPrecodingCNN, ELM} (nested) \\
		\hline
		6D & Hybrid precoding, wideband SU-MIMO \cite{learnOFDM, yuweiHybridOFDM} & Hybrid precoding, wideband MU-MIMO \cite{MU-MIMO-OFDM} (nested)\\
		\hline\hline
	\end{tabular}
	\vspace{-4mm}
\end{table*}

\subsubsection{2D-GNN for Learning Coordinated Beamforming Policy}
Consider an interference system with $B$ BSs, where each BS with $N_t$ antennas serves $K$ users in each cell.
%When a BS transmits to a user associated to it, other BSs should not generate interference to the user, as shown in Fig. \ref{fig:fig-example} (d). The difference between the CoMP-CB problem and the single-cell MU-MISO precoding problem lies in the power constraints of each BS \cite{WMMSE}.
%The sum-rate problem is a permutable problem.

There are two nested sets in the coordinated beamforming problem:
the user-set with $BK$ elements, the antenna-set with $BN_t$ elements, similar to the problem illustrated in Fig. \ref{fig:fig-example}(d). The permutations of all users and all antennas are partially dependent, which are respectively denoted as ${\bf \Omega}^T_1\triangleq({\bf \Pi}_3^T\otimes {\bf I}_K){\rm diag}({\bf \Pi}_{1,1}^T,\cdots,{\bf \Pi}_{1,B}^T)$ and ${\bf \Omega}^T_2\triangleq({\bf \Pi}_3^T\otimes {\bf I}_{N_t}){\rm diag}({\bf \Pi}_{2,1}^T,\cdots,{\bf \Pi}_{2,B}^T)$, where ${\bf \Pi}_3$  is the permutation of BSs, ${\bf \Pi}_{1,b}$ and ${\bf \Pi}_{2,b}$ are respectively the permutation of users and antennas in the $b$-th cell. The coordinated beamforming policy ${\bf W}_{BB}=f({\bf H})$ satisfies a \emph{nested joint permutation property}: ${\bf \Omega}^T_1{\bf W}_{BB}^T{\bf \Omega}_2=f({\bf \Omega}_1^T{\bf H}{\bf \Omega}_2)$.

In the established graph, users and antennas are vertices, ${\bf H}\in \mathbb{C}^{ BK\times BN_t}$ and ${\bf W}_{BB}\in \mathbb{C}^{ BN_t\times BK}$ are the edge-features between user and AN vertices. A 2D-GNN can be used to learn this policy, the input layer and output layer are the same as those in  section \ref{subsec:inde_nest}.
%to those of problems with dependent and arbitrary permutations, due to the same dimensions of features and hidden representations ${\bf X}_l\in \mathbb{R}^{C_l\times BK\times BN_t}$.
In the update layers, the structure of ${\bf P}_l$ in ${\rm vec}({{\bf X}_{l+1}}) = \sigma ({\bf P}_l{\rm vec}({{\bf X}_l}))$  with ${\bf X}_l\in \mathbb{R}^{C_l\times BK\times BN_t}$ can be designed by first using the method in \cite{equ} and then setting the weights for non-adjacent edges as zero.

{\bf Remark 5:} $M$-dimensional GNNs are designed to avoid information loss meanwhile exploit the $M$ possible permutations when learning the policies from $M$-set problems. The information loss is avoided by updating  hidden representations ${\bf X}_l \in \mathbb{R}^{C_l\times N_1\times N_2 \cdots \times N_M}$ in the $M$-dimensional feature space.
All permutations are exploited by designing each layer for satisfying the properties in Proposition 1 and constructing the graph with $M$ types of vertices. The hypothesis space of a $(M-1)$-dimensional GNN contains the hypothesis space of a $M$-dimensional GNN, but $(M-1)$-dimensional GNN is not a special case of $M$-dimensional GNN.

%TABLE
In Table \ref{table:problems}, we list some wireless policies that can be learned by the proposed MD-GNN framework, where the precoding without ``hybrid'' means baseband precoding and the multi-antenna systems without ``wideband'' means narrow-band multi-antenna systems.

\section{Practical Issues of using the MD-GNNs}\label{practical}
In this section, we consider two issues when using MD-GNNs for learning wireless policies.
\subsection{Are Input Samples Permutable?}\label{sec:sample}\
GNNs are efficient for learning permutable functions. Yet in practice, the permutability of a policy depends on the distribution of environment parameters. For example, when learning the policy from problem $P1$, $K!N_t!$ permuted samples are possible to be gathered or generated if the channels are independent and identically
distributed (i.i.d.).  However, the channels in multi-antenna systems are often spatially correlated due to the tightly packed antenna arrays and the sparsely scattered propagation environments. Hence, a natural question is: are there $K!N_t!$ permuted versions for each channel matrix in the input feature space of dimension ${K\times N_t}$?
If the answer is not, then the precoding policy is not fully permutable, which cannot yield $K!N_t!$ equivalent feasible solutions. As a consequence, the gain of GNNs  in terms of learning efficiency over the FNNs without embedding any prior will be lower. Therefore, we are interested in the permutability of environment parameters, specifically wireless channels.

We answer this question by taking  precoding problem in a MISO-OFDM system as an example, where the permutability of channel tensors depends on channel correlation.

{\bf Users:}
The permutability of a channel tensor along the user dimension depends on the correlation of the channels among users. For $K$ users not closely located, their channel matrices are independent, and there exists $K!$ permuted version of one channel tensor in the feature space.
%one sample can represent $K!$ permuted version samples.

%\subsection{Antennas}\label{subsubsec:sample-antenna}
{\bf Antennas:}
To understand the impact of spatial correlation on the permutability of the channels of a single user, we first consider the narrow-band Saleh-Valenzuela (SV) channel model that can capture the mathematical structure of mmWave channels. For simplicity, we consider uniform linear antenna array. The channel from the BS to one user can be modeled as ${\bf h} = \sqrt{\frac{N_t}{N_{cl}N_{ray}}}\sum_{i=1}^{N_{cl}}\sum_{j=1}^{N_{ray}}\alpha_{i,j} {\bf f}(\theta_{i,j}) \in \mathbb{C}^{N_t}$\cite{MO},
%\begin{equation*}
%{\bf h} = \sqrt{\frac{N_t}{N_{cl}N_{ray}}}\sum_{i=1}^{N_{cl}}\sum_{j=1}^{N_{ray}}\alpha_{i,j} {\bf f}(\theta_{i,j}) \in \mathbb{C}^{N_t},
%\end{equation*}
where $N_{cl}$ is the number of scattering clusters, $N_{ray}$ is the number of scattering rays, $\alpha_{i,j}$ is a complex gain, and ${\bf f}(\theta_{i,j})=\frac{1}{\sqrt{N_t}}[1, e^{j\frac{2\pi d}{\lambda}\sin \theta_{i,j}},\cdots,e^{j(N_t-1)\frac{2\pi d}{\lambda}\sin \theta_{i,j}}]^T$ is the array response with angle of departure (AoD) $\theta_{i,j}$, wavelength $\lambda$, and antenna spacing $d$.
%\begin{equation*}
%{\bf f}(\theta_{i,j})=\frac{1}{\sqrt{N_t}}[1, e^{j\frac{2\pi d}{\lambda}\sin \theta_{i,j}},\cdots,e^{j(N_t-1)\frac{2\pi d}{\lambda}\sin \theta_{i,j}}]^T,
%\end{equation*}

Consider two channel vectors ${\bf h}$ and ${\bf h}^{\prime}= \sqrt{\frac{N_t}{N_{cl}N_{ray}}}\sum_{i=1}^{N_{cl}}\sum_{j=1}^{N_{ray}}\alpha_{i,j}^\prime {\bf f}\left( \theta_{i,j}^\prime \right)$. If we can find $\alpha_{i,j}$, $\alpha'_{i,j}$, $\omega_{i,j}$, and $\omega'_{i,j}$ that satisfy ${\bf h}^\prime = {\bf \Pi}_2^T {\bf h}$, where $\omega_{i,j} \triangleq \exp(j2\pi d \sin \theta_{i,j}/\lambda)$, and ${\bf \Pi}_2$ is the permutation of antennas,
then there exist channel vectors  in the feature space that are the permuted versions of ${\bf h}$. Upon substituting the array response, ${\bf h}^\prime = {\bf \Pi}_2^T {\bf h}$ can be rewritten as a group of scalar equations as\vspace{-2mm}
\begin{equation*}
\vspace{-2mm}
\sum_{i=1}^{N_{cl}}\sum_{j=1}^{N_{ray}}( \alpha_{i,j}^\prime {\omega_{i,j}^\prime}^{n-1} - \alpha_{i,j} \omega_{i,j}^{\pi_2(n) -1}) = 0, n=1,\cdots,N_t,
\end{equation*}
which can be considered as $N_t$ linear equations with $2N_{cl}N_{ray}$ unknowns $\alpha_{i,j}^\prime, \alpha_{i,j}, i=1,\cdots,N_{cl}, j=1,\cdots,N_{ray}$, where ${\omega_{i,j}^\prime}^{n-1}$ and $\omega_{i,j}^{\pi_2(n) -1}$ are coefficients of the equations.

If $N_{cl}N_{ray} \ge N_t/2$, then the number of equations is less than the number of unknowns, and there always exist solutions regardless of the value of ${\bf\Pi}_2$. This implies there always exist $N_t!$ permutable samples in the feature space, each can be obtained by permuting another one.

If $N_{cl}N_{ray} < N_t/2$, the equations only have solutions for specific permutations. For example, if $1/\omega_{i,j}= \omega_{i,j}^\prime$ then there exists a solution $\alpha_{i,j} \omega_{i,j}^{N_t-1}=\alpha_{i,j}^\prime$  with permutation $\pi_2(n) = N_t + 1 - n $; and
%This special permutation is reversing, as shown in Fig. \ref{fig:fig-array}. The other one is that,
if $\omega_{i,j} = \omega_{i,j}^\prime=2l\pi/N_t$ then there exists $N_t$ solutions $\alpha_{i,j}\omega_{i,j}^{(1+k){\rm mod}N_t}=\alpha_{i,j}^\prime$ with permutations $\pi_2(n) = (n+k){\rm mod}N_t + 1, l,k \in \mathbb{Z}$, which can be  further permuted by $\pi_2(n) = N_t + 1 - n $, where ${\rm mod}$ stands for the modulo operation.
%This special permutation is circular shifting (rotating).
Hence, there exist $2N_t$ permutable samples regardless of the values of $N_{cl}$ and $N_{ray}$,
i.e., one channel vector $\bf h$ has at least $2N_t$ permuted versions in the feature space.
%With more scattering clusters and rays, more permutations appear, until all $N_t!$ permutations exist.
%are  certainly included when $N_{cl}N_{ray} \ge N_t/2$.

%\begin{figure}[htb]
%	\centering
%	\vspace{0mm}
%	\includegraphics[width=0.3\linewidth]{Fig-array}
%	\vspace{-4mm}
%\caption{Two special permutations exist in multi-antenna channels even without multipath ($N_{cl}N_{ray}=1$). All the four circles are the possible channels with antennas ordered. The locations of red numbers represent the phases of elements, they are equally spaced due to the array responses. ``1'' is not on zero degree because of random complex gain $\alpha_{1,1}$. There are $N_t$ different circular shifting. Thus, taking reversing into account, there exist $2N_t$ permutations regardless the values of $N_{cl}$ and $N_{ray}$.}
%	\vspace{-4mm}
%	\label{fig:fig-array}
%\end{figure}

%\subsection{Subcarriers}\label{subsebsec:subcarrier}
{\bf Subcarriers:} To show the impact of frequency correlation on the permutability of the
channels in subcarriers, we consider
a tap delay-$d$ channel model in wideband mmWave systems. The channel from a BS with uniform linear antenna array to a user can be modeled as $\hat{{\bf h}}^d = \sqrt{\frac{N_t}{N_{cl}N_{ray}}}\sum_{i=1}^{N_{cl}}\sum_{j=1}^{N_{ray}}\alpha_{i,j} {\bf f}(\theta_{i,j})p(dT_s\! -\! \tau_{i,j})$\cite{learnOFDM},
%\begin{equation*}
%\hat{{\bf h}}[d] = \sqrt{\frac{N_t}{N_{cl}N_{ray}}}\sum_{i=1}^{N_{cl}}\sum_{j=1}^{N_{ray}}\alpha_{i,j} {\bf f}(\theta_{i,j})p(dT_s -\tau_{i,j}),
%\end{equation*}
where $T_s$ is the symbol duration, $\tau_{i,j}$ is the delay of the $j$-th ray in the $i$-th cluster, and $p(\cdot)$ is a pulse shaping function. Denote the number of taps as $D$ and assume that the length of cyclic prefix is larger than $DT_s$. After a $M$-point discrete Fourier transform, the channel at the $m$-th subcarrier is ${\bf h}^m = \sum^{D-1}_{d=0} \hat{{\bf h}}^d e^{-j\frac{2\pi m}{M}d}$.
%\begin{equation*}
%{\bf h}[m] = \sum^{D-1}_{d=0} \hat{{\bf h}}[d]e^{-j\frac{2\pi m}{M}d}.
%\end{equation*}

The permutability of $M$ subcarriers depends on $D$. If $D=1$ (i.e., flat fading), then the channels of all subcarriers are identical and hence can be permuted arbitrarily.
To analyze the permutability of the channel with $D>1$, we examine when the equations ${\bf h}'^m= {\bf h}^{\pi_4(m)},m=1,\cdots,M$ have solutions, which have $2D$ unknowns: $\hat{{\bf h}}^1,\cdots,\hat{{\bf h}}^D,$ $\hat{{\bf h}}'^1,\cdots,\hat{{\bf h}}'^D$, where $\pi_4(m)$ represents the permutation of subcarriers. If $D\ge M/2$, then the number of equations is less than the number of unknowns. Hence,
there always exist solutions, which means that $M!$ permuted samples are available. If $M/2 > D\ge 2$, $M$ specific permutations $\pi_4(m)=(m+k){\rm mod}M+1, k\in\mathbb{Z}$ always make the equations solvable.%, because ${\bf h}^m$ and ${\bf h}'^m$ are two sums of unknowns with specific phase.
%Besides, $e^{-j2\pi d/M}$ is always negative phase, the reversing permutation does not exist. As taps increase, other permutations may emerge.

In summary, even in the worst case where $N_{cl}N_{ray}=1 \ll N_t$, $D=2\ll M$, the channel tensors are permutable, and each channel has $2K!N_tM$ permuted versions in the feature space.
%All the $K!N_t!M!$ permutations may exist, depending on the propagation environments.

\subsection{Tradeoff Between Training, Inference, and Design Complexities}\label{reducingdim}
Training complexity includes sample and space complexities  (i.e., the minimal numbers of training samples and trainable parameters required by a deep neural network (DNN) to achieve a given performance)  and training time. Since a DNN is trained offline, the training time is of less concern. Since learning-based solution is used for reducing inference time and gathering samples from real environments is expensive, we focus on the time complexity for inference, sample complexity, and space complexity that is the same for both training and inference.

To show the potential of GNNs in exploiting permutation prior, the number of vertex types of a graph should be equal to the number of sets in a problem. To avoid information loss, the order of the hidden representation tensor ${\bf X}_l$ in the MD-GNNs should be equal to the number of vertex types plus one.
If we purposely harness less permutations, we can construct a graph with less type of vertices, and then  ${\bf X}_l$ will be with lower order. As a result, the MD-GNN may become faster for inference, but needs higher sample complexity due to enlarged hypothesis space.
%To achieve a tradeoff between fast inference and learning performance

Take problem $P1$ as an example. If the permutation of RF chains is ignored, then
%the two output features can be regarded as ${\bf W}_{RF}\in \mathbb{R}^{2N_s\times N_t}$ and ${\bf W}_{BB}\in \mathbb{R}^{2N_s\times K }$, and
the hybrid precoding policy can be learned with a 2D-GNN over a graph only with user-vertices and AN-vertices.
In the input layer of the 2D-GNN, it is no need to increase dimension. The update layers  are illustrated in Fig. \ref{fig:fig-update}(a) with ${\bf X}_l\in \mathbb{R}^{C_l\times K \times N_t}$. In the output layer, there are $C_L=4N_s$ ``channels'' to represent the real and imaginary parts of ${\bf W}_{RF}$ and ${\bf W}_{BB}$.
% in different RF chains.
The hypothesis space of this 2D-GNN is larger than that of the 3D-GNN designed in section \ref{subsec:3dgnn}.
%The  2D-GNN learns a permutable function with the permutations of users and antennas, whereas the 3D-GNN learns a permutable function with one more permutation of RF chains. Hence, the
The 3D-GNN leverages the prior that RF chains are permutable and the two precoding matrices should be permuted simultaneously by ${\bf \Pi}_3$ as shown in \eqref{eq:taskfucPE2}, but the 2D-GNN has to learn the knowledge from samples. Similarly, the hybrid precoding policy can also be learned with a 1D-GNN over a graph only with AN-vertices.

%\subsection{Analysis}
%Although a higher dimensional GNN can exploit the prior knowledge of a permutable problem, it is with higher computational complexity. We can reduce the dimension of GNNs for real-time applications. Similar to the above 2D-GNN, the conversion need reducing the number of dimensions and increasing the number of channels of outputs, but without effect on input or output features.

However, the 3D-GNN is with higher time complexity for inference, measured in floating point operations (FLOPs). To see this, we compare the number of FLOPs and its order of magnitude of the 3D-GNN with 1D-GNN, and 2D-GNN.

For comparison, we also provide the FOLPs of commonly used CNN. For the CNN  with $C_l$ and $C_{l+1}$ ``channels'' in the $l$-th and $(l+1)$-th layers,
%for each convolution kernel with size $C_l S_H S_W$,
 $C_l S_H S_W$ multiplications and $C_l S_H S_W-1$ additions are required, where $S_H\times S_W$ is the size of convolutional kernel. These multiplications and additions need to be used $C_{l+1}KN_t$ times in the $(l+1)$-th layer. Thus, the FLOPs to compute the hidden representation in the $(l+1)$th layer is $(2C_lS_HS_W-1)C_{l+1}KN_t$.

For the MD-GNNs, we consider the update layer with parameter sharing  in \eqref{PS-P} and use non-zero weights only for aggregating adjacent edges. Implementing the update layer ${\rm vec}({{\bf X}_{l+1}}) = \sigma ({\bf P}_l{\rm vec}({{\bf X}_l}))$ needs to sum some particular elements in ${{\bf X}_l}$ and then multiply the summation by a weight only once. When we sum and weight particular elements to update an element (e.g., the red element as illustrated in Fig. 4), the summation can be reused for updating different elements. We explain this in detail in the following.

For 1D-GNN, ${{\bf X}_l}=[{\bf x}_{l,1},\cdots, {\bf x}_{l,N_t}]\in \mathbb{R}^{C_l\times N_t}$ includes the hidden representations of all antenna-vertices. The representation of the $n$-th vertex ${\bf x}_{l+1,n}$ is updated by ${\bf x}_{l+1,n} = \sigma({\bf P}_{l,1}{\bf x}_{l,n}+{\bf P}_{l,2}\sum_{m=1,m\neq n}^{N_t}{\bf x}_{l,m})$, where ${\bf P}_{l,1},{\bf P}_{l,2}\in \mathbb{R}^{C_{l+1}\times C_l}$ include $2C_lC_{l+1}$ weights. The update equation can also be written as ${\bf x}_{l+1,n} = \sigma({\bf P}'_{l,1}{\bf x}_{l,n}+{\bf P}_{l,2}\sum_{m=1}^{N_t}{\bf x}_{l,m})$, where ${\bf P}'_{l,1}$ includes $C_lC_{l+1}$ weights. The summation $\sum_{m=1}^{N_t}{\bf x}_{l,m}$ needs $C_l(N_t-1)$ additions, and the value of the summation can be reused. Then, both ${\bf P}'_{l,1}{\bf x}_{l,n}$ and ${\bf P}_{l,2}\sum_{m=1}^{N_t}{\bf x}_{l,m}$ are the products of a matrix and a vector, each needs $C_{l+1}C_l$ multiplications and  $C_{l+1}(C_l-1)$ additions for computing. ${\bf P}'_{l,1}{\bf x}_{l,n}$ has to be computed $N_t$ times for antenna vertices. Finally, adding ${\bf P}'_{l,1}{\bf x}_{l,n}$ and ${\bf P}_{l,2}\sum_{m=1}^{N_t}{\bf x}_{l,m}$ needs $C_{l+1}$ additions for each antenna vertex. After considering all additions and multiplications, $(2C_l-1)C_{l+1}(N_t+1)+C_l(N_t-1)+C_{l+1}N_t$ FLOPs are required for the updating in the $(l+1)$th layer.

For 2D-GNN, ${{\bf X}_l}\in \mathbb{R}^{C_l\times K\times N_t}$ includes the hidden representations of all edges between antenna- and user-vertices. The  representation of the edge between the $k$-th user and the $n$-th antennas ${\bf x}_{l+1,k,n}\in  \mathbb{R}^{C_{l+1}}$ is updated by ${\bf x}_{l+1,k,n} = \sigma({\bf P}_{l,1,1}{\bf x}_{l,k,n}+{\bf P}_{l,2,1}\sum_{i=1,i\neq k}^{K}{\bf x}_{l,i,n}+{\bf P}_{l,1,2}\sum_{m=1,m\neq n}^{N_t}{\bf x}_{l,k,m})$, where ${\bf P}_{l,1,1},{\bf P}_{l,2,1},{\bf P}_{l,1,2}\in \mathbb{R}^{C_{l+1}\times C_l}$ include $3C_lC_{l+1}$ weights. Computing the two summations $\sum_{i=1}^{K}{\bf x}_{l,i,n}$ and $\sum_{m=1}^{N_t}{\bf x}_{l,k,m}$ needs $C_l(K-1)N_t + C_lK(N_t-1)$ additions. Computing the three products of a matrix and a vector need $C_{l+1}(2C_l-1)(KN_t+K+N_t)$ FLOPs for every antenna- and user-vertex. Computing the summation of the three terms in the update equation needs $2C_{l+1}KN_t$ additions. In summary, $(2C_l -1)C_{l+1}(KN_t+K+N_t) +C_l[K(N_t-1)\! +\!(K-1)N_t]+2C_{l+1}KN_t$ FLOPs are required for the update in the $(l+1)$th layer.

Using similar derivations (omitted due to the space limitation), the number of FLOPs required by updating one layer of the 3D-GNN can be computed as $C_{l+1}(2C_l- 1)(KN_tN_s + KN_t + KN_s + N_tN_s) + C_l[KN_t(N_s - 1) + K(N_t - 1)N_s + (K -1)N_tN_s] + 3C_{l+1}KN_tN_s$.
%For 3D-GNN, ${{\bf X}_l}\in \mathbb{R}^{C_l\times K\times N_t\times N_s}$ includes the hidden representations of all hyper-edges between antenna-vertices, user-vertices, and RF-chain-vertices. The  representation of the hyper-edge between the $k$-th user, the $n$-th antennas, and the $l$-th RF chain ${\bf x}_{l+1,k,n,l}\in  \mathbb{R}^{C_{l+1}}$ is updated by ${\bf x}_{l+1,k,n,l} = \sigma({\bf P}_{l,1,1,1}{\bf x}_{l,k,n,l}+{\bf P}_{l,2,1,1}\sum_{i=1,i\neq k}^{K}{\bf x}_{l,i,n,l}+{\bf P}_{l,1,2,1}\sum_{m=1,m\neq n}^{N_t}{\bf x}_{l,k,m,l}+{\bf P}_{l,1,1,2}\sum_{j=1,j\neq l}^{N_s}  {\bf x}_{l,k,n,j})$,  where ${\bf P}_{l,1,1,1}, {\bf P}_{l,2,1,1}, {\bf P}_{l,1,2,1}, {\bf P}_{l,1,1,2} \in \mathbb{R}^{C_{l+1}\times C_l}$ include $4C_lC_{l+1}$ weights.
%%These weight matrices are same to the scalars in equation (7) of the manuscript, because we let $C_l=C_{l+1}=1$ for a concise derivation in manuscript.
%The three summations need $C_l[KN_t(N_s - 1) + K(N_t - 1)N_s + (K -1)N_tN_s]$ additions. The four products need $C_{l+1}(2C_l- 1)(KN_tN_s + KN_t + KN_s + N_tN_s)$ FLOPs. Adding the four terms in the update equation needs $3C_{l+1}KN_tN_s$ additions. In summary, $C_{l+1}(2C_l- 1)(KN_tN_s + KN_t + KN_s + N_tN_s) + C_l[KN_t(N_s - 1) + K(N_t - 1)N_s + (K -1)N_tN_s] + 3C_{l+1}KN_tN_s$ FLOPs are required between two layers.

The order of magnitude of the number of FLOPs for inference and the number of trainable parameters are listed in Table \ref{table:compute}. It is shown that space complexities of all the DNNs are of the same order of magnitude, but time complexities of MD-GNNs grow with the dimensions.
%3D-GNN has the highest inference complexity and the lowest sample complexity (to be validated in the next section).

\begin{table}[htb!]
	\setlength\tabcolsep{2pt}
	\centering
	\caption{Time complexity for inference and space complexity}\label{table:compute}
	\vspace{-2mm}
	\footnotesize
	\renewcommand\arraystretch{1}
	\begin{tabular}{c|c|c}
		\hline\hline
		DNNs & Order of magnitude of FLOPs & Number of trainable parameters\\
		\hline
		CNN & $O(S_HS_WC_lC_{l+1}KN_t)$ & $S_HS_WC_lC_{l+1}$\\
		\hline
		1D-GNN & $O(C_lC_{l+1}N_t)$& $2C_lC_{l+1}$\\
		\hline
		2D-GNN & $O(C_lC_{l+1}KN_t)$ & $3C_lC_{l+1}$\\
		\hline
		3D-GNN & $O(C_lC_{l+1}KN_tN_s)$ &  $4C_lC_{l+1}$\\
		\hline\hline
	\end{tabular}
	\vspace{-2mm}
\end{table}

This suggests that we can make the following tradeoff.

If one prefers a fast GNN without increasing sample complexity significantly, the permutation of the following sets can be deliberately unharnessed:
a) the sets that are not associated with any input features  (e.g., RF chain-set), because ignoring these sets has little impact on the sample complexity;
b) the sets with few elements  (e.g., user-set if users are much fewer than antennas), because the permutations for the elements in these sets are relatively few;
c) the sets whose elements are with specific distributions  (e.g., the subcarrier-set in frequency-selective channel with a small value of $D$), because the permutable channel samples are relatively few.

This also allows us to trade off the design complexity with sample complexity. For example, we can simply use 2D-GNNs for all the problems with more than two sets at the cost of higher sample complexity, without the need to re-design complicated input, update, and output layers. One can also simply use a 1D-GNN for all permutable problems.

\section{Simulation Results}\label{sec:result}
In this section, we take the narrow-band and wideband hybrid precoding policies in mmWave MISO systems as examples to evaluate the performance of the proposed MD-GNNs. In particular, we learn the policy $({\bf W}_{RF}, {\bf W}_{BB}) = f({\bf H})$.

\subsection{Learning Narrow-band Hybrid Precoding Policy from Problem $P1$}\label{subsec:result-hybrid-precoding}
A BS with 64 antennas and six RF chains serves three users. The signal-to-noise ratio (SNR), $P_{tot}/{\sigma ^2}$, is 10 dB. Channels are generated with the SV model with uniform linear array, where the AoDs are uniformly selected from $[0, 2\pi)$ with angular spread of 10 degrees in each cluster, $d = \lambda/2$, $N_{cl}=4$, $N_{ray}=5$, and $\alpha_{i,j} \sim \mathcal{CN}(0,1)$ \cite{MO}.

We generate 500,000 channel samples to train and 10,000 samples to test the DNNs. We apply ReLU as activation function and Adam as optimizer. The batch-size is 500, and batch-normalization is used. The initial learning rate is 0.001. We consider unsupervised learning, where the loss function is the negative sum-rate (we set $\beta_k=1, k=1,\cdots, K$ in the simulation) in \eqref{objective1} averaged over a batch of training samples.

We compare the proposed 3D-GNN with CNN \cite{HybridPrecodingCNN,BF2021}, FNN \cite{DNN2019},
as well as the 2D-GNN and 1D-GNN in section \ref {reducingdim}, where the CNN is composed of convolutional layers with $3\times3$ convolutional kernel followed by a fully-connected layer.
We also compare with other three GNNs that satisfy the three-set permutation property. The first is a three-set GNN (TGNN) proposed in \cite{LSJ}, which can also avoid the information loss when updating the hidden representation of edges. The second is a revised version of the edge-GNN in \cite{ZBC} that updates edge representations for analog and baseband precoding
matrices alternately. The third is a revised version of a vertex-GNN designed in \cite{Korea} for optimizing baseband precoding indirectly by leveraging the structure of the sum-rate maximal optimal precoding matrix. The revised vertex-GNN updates the hidden representations of the three types of vertices in each hidden layer, where six three-layer FNNs are used for processing the information from other types of adjacent vertices and other types of adjacent edges before sum pooling, three three-layer FNNs are used in the combination for each type of vertex, and another two FNNs are used in the output layer for mapping the hidden representation to the two precoding matrices.

This setup is used unless otherwise specified, and all the DNNs have been fine-tuned.
%by combining itself and . In the vertex-GNN,
%Information loss may occur while edge information is mixed, although non-linear functions try to avoid this.

We simulate the following numerical algorithms, the MO algorithm \cite{MO}, the PEM algorithm \cite{MO}, and the OMP algorithm \cite{OMP}, for comparison.

\begin{figure}[htb!]
	\centering
	\subfigure[$N_t=16$]{
		\begin{minipage}[c]{0.8\linewidth}
			\centering
			\includegraphics[width=1\linewidth]{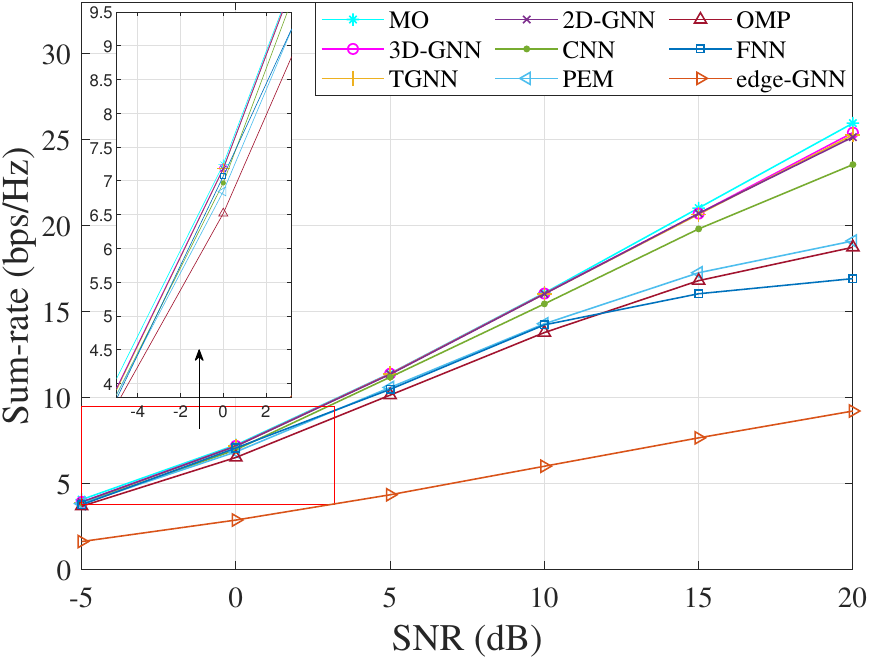}
		\end{minipage}
	}\\	\vspace{-2mm}
	\subfigure[$N_t=64$]{
		\begin{minipage}[c]{0.8\linewidth}
			\centering
			\includegraphics[width=1\linewidth]{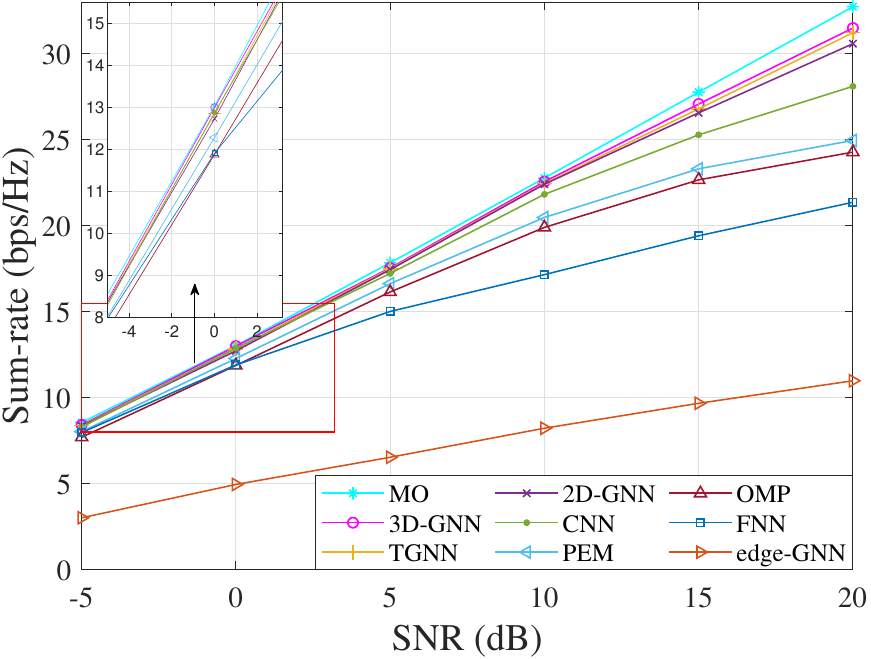}
		\end{minipage}
	}
	\vspace{-2mm}
	\caption{Impact of SNR, $K=3$, $N_s=6$.}
	\label{fig:fig-SNR}
	\vspace{-2mm}
\end{figure}

In Fig. \ref{fig:fig-SNR}, we show the impact of SNR. It is shown that GNNs are superior to CNN and FNN when SNR is high, and 3D-GNN achieves a slightly higher sum-rate than 2D-GNN when $N_t$ is large. The revised version of the edge-GNN in \cite{ZBC} performs very poor due to the information loss. To provide a clean figure, we do not show the results of the 1D-GNN and vertex-GNN here. Since TGNN achieves comparable performance with 3D-GNN but is much harder to train and is with much higher inference complexity, we do not compare with it again in the sequel.

Since the MO algorithm achieves the highest sum-rate, we use it as the benchmark in the following. In particular, we compare the learning performance  achieved by the DNNs in terms of sum-rate relative to the MO algorithm (in percentage).

In Table \ref{table:Nt}, we show the impact of the number of antennas. We can see that the policies learned by GNNs perform close to the MO algorithm. The performance  gains of GNNs over FNN and CNN grow with the values of $N_t$. Despite that both the 3D-GNN and the vertex-GNN exploit the three-set permutation property, the 3D-GNN provides slightly higher sum-rate. This is because the vertex-GNN may still lose information during aggregating at vertices, although the six FNNs for processing are with 256 neurons (also called ``channels'') in their output layers.
%when 128 ``channels'' are used for processing information in the setting with $N_t=8$ or 16 and 256 ``channels'' are used for the setting with $N_t=36$ or 64 (i.e., the output dimensions of the six FNNs  for processing are 128 and 256, respectively). %In fact, it is unable to know  when the output dimensions of the six FNNs are sufficient large to avoid the information loss.
%Since 1D-GNN and vertex-GNN perform close to 2D-GNN, we no longer provide their sum rates in the following.
\begin{table}[htb!]
	\vspace{-4mm}
	\setlength\tabcolsep{2pt}
	\centering
	\caption{Sum-rate versus number of antennas ($K=3$,$N_s=6$,SNR=10$ \rm dB$)}\label{table:Nt}
	\vspace{-2mm}
	\footnotesize
	\renewcommand\arraystretch{1}
	\begin{tabular}{c|c|c|c|c|c|c|c}
		\hline\hline
		\multirow{2}{*}{$N_t$} & \multirow{2}{*}{\makecell[c]{MO \\ (bps/Hz)}} & \multicolumn{6}{c}{Sum-rate relative to MO}\\
		\cline{3-8}
		 &  & FNN & CNN &Vertex-GNN & 1D-GNN & 2D-GNN &3D-GNN \\
		\hline
		8&  12.39& 95.48\%& 96.77\%& 98.78\%& 98.03\%& 98.95\%& 99.11\% \\
		\hline
		16& 16.09& 88.44\%& 96.02\%& 98.61\%& 99.62\%& 99.75\%& 99.80\% \\
		\hline
		36&	20.10& 82.94\%&	96.17\%& 98.23\%& 98.81\%& 98.71\%& 99.25\% \\
		\hline
		64&	22.77& 75.32\%&	95.83\%& 98.14\%& 98.29\%& 98.43\%& 99.08\% \\
		\hline\hline
	\end{tabular}
	\vspace{-2mm}
\end{table}

In Table \ref{table:K}, we show the impact of the number of users. When $K$ increases, the performance of all DNNs decreases, but the decreasing rate of the GNNs is smaller. This is especially true for the A2D-GNN mentioned in Remark 1, whose performance degrades slightly with more users.

\begin{table}[htb!]
	\vspace{-2mm}
	\setlength\tabcolsep{2pt}
	\centering
	\caption{Sum-rate versus number of Users ($N_t=64$, SNR=10$\rm dB$)}\label{table:K}
	\vspace{-2mm}
	\footnotesize
	
	\renewcommand\arraystretch{1}
	\begin{tabular}{c|c|c|c|c|c|c|c}
		\hline\hline
		\multirow{2}{*}{$K$} &\multirow{2}{*}{$N_s$} & \multirow{2}{*}{\makecell[c]{MO \\ (bps/Hz)}} & \multicolumn{4}{c}{Sum-rate relative to MO}\\
		\cline{4-8}
		 & &  &  FNN & CNN & 2D-GNN &3D-GNN &A2D-GNN\\
		\hline
		3&	6&	22.77& 75.32\%&	95.83\%& 98.43\%& 99.08\% & 99.83\%\\
		\hline
		4&	8&	28.60&  63.64\%&  88.71\%&  95.31\%&  96.47\% & 99.80\%\\
		\hline
		6&	12&	38.99&  52.06\%&  83.66\%&  91.25\%&  93.40\% & 99.70\%\\
		\hline
		10&	16&	56.35&  36.95\%&  69.44\%&  86.12\%&  90.09\% & 99.38\%\\
		\hline
		16&	16&	68.28&  35.41\%&  57.78\%&  81.14\%&  84.52\% & 98.71\%\\
		\hline\hline
	\end{tabular}\vspace{-2mm}
\end{table}

In Table \ref{table:Ncl}, we show the impact of spatial correlated channels. We also train and test the DNNs with the DeepMIMO dataset \cite{deepmimo}, which considers an outdoor scenario over 28 GHz with a strong line-of-sight ray. As expected, GNNs perform better for the channels with more scattering rays, which validates the analysis in section \ref{sec:sample}. The results for the 1D-GNN, A2D-GNN, and vertex-GNN are similar with 2D-GNN, hence are not provided for conciseness.

\begin{table}[htb!]
\setlength\tabcolsep{2pt}
	\vspace{-2mm}
	\centering
	\vspace{-2mm}
	\caption{Sum-rate versus Scattering Rays ($K=3$, $N_s=6$, $N_t=16$, SNR=10$\rm dB$)}\label{table:Ncl}
	\vspace{-2mm}
	\footnotesize
	
	\renewcommand\arraystretch{1}
	\begin{tabular}{c|c|c|c|c|c|c}
		\hline\hline
		\multirow{2}{*}{$N_{cl}$} &\multirow{2}{*}{$N_{ray}$} & \multirow{2}{*}{\makecell[c]{MO \\ (bps/Hz)}} & \multicolumn{4}{c}{Sum-rate relative to MO}\\
		\cline{4-7}
		& & & FNN & CNN & 2D-GNN &3D-GNN\\
		\hline
		1&	3&	15.14&	91.22\%& 95.71\%&	97.56\%& 97.82\%\\
		\hline
		2&	3&	15.73&	89.83\%& 96.31\%&	98.03\%& 98.35\%\\
		\hline
		4&	5&	16.09&	88.44\%& 96.02\%&	99.75\%& 99.80\%\\
		\hline
		8&	10&	16.27&	89.80\%& 96.37\%&	99.74\%& 99.78\%\\
		\hline
		\multicolumn{2}{c|}{DeepMIMO} &13.90& 94.79\%& 95.12\%& 97.71\%& 97.55\%\\
		\hline\hline
	\end{tabular}
	\vspace{-2mm}
\end{table}

In Tables \ref{table:scale} and \ref{table:samples}, we provide space complexity and sample complexity of the DNNs except FNN to achieve 95\% sum-rate of MO algorithm. The results of FNN are obtained when it achieves its best performance (lower than 95\%) with all the 500,000 training samples.

We can see that the space complexities of FNN and CNN are much higher than the GNNs, due to the larger hypothesis space. Since the vertex-GNN employs $3$ (layers) $\times 9+2=29$ FNNs for processing, combination, and mapping the vertex representation to precoders, its space complexity (after fine-tuning the 29 FNNs) is much higher than the 3D-GNN. The  space complexity of 1D-GNN exceeds 2D-GNN, and both exceed 3D-GNN, which seems inconsistent with the results in Table \ref{table:compute}. This is because 1D-GNN needs more ``channels'' than 2D-GNN, and both need more  ``channels'' than 3D-GNN.
The sample complexities of the proposed GNNs are much fewer than FNN and CNN, and decrease with the increase of $N_t$ except 64 antennas (because not all of the $64!$ permutations of the samples along the antenna dimension exist in the feature space). To obtain a deeper insight into GNNs with different dimensions, we also provide the sample complexities of a 2D'-GNN that only considers permutations of antennas and RF chains. The 2D-GNN and 3D-GNN can harness $K!N_t!$ permutations, whereas 1D-GNN and 2D'-GNN can only harness  $N_t!$ permutations.  The less sample demand of 3D-GNN and 2D-GNN validates the analysis in section \ref{reducingdim}.
%It can be seen that both the hypothesis space and the permutability of samples affect the sample complexity.
The much higher sample complexity of vertex-GNN than 3D-GNN stems from the need to learn injective aggregation functions \cite{howPowerful}.

\begin{table}[htb!]
	\vspace{-2mm}
	\centering
	\setlength\tabcolsep{0pt}
	\vspace{-2mm}
	\caption{Space complexity	($K=3$, $N_s=6$, $N_t=16$, SNR=10$\rm dB$)}\label{table:scale}
	\vspace{-2mm}
	\footnotesize
	\renewcommand\arraystretch{1}
	\begin{tabular}{c|c|c|c|c|c|c}
		\hline\hline
		~ &	FNN &	CNN &Vertex-GNN &1D-GNN & 2D-GNN  & 3D-GNN \\
		\hline
		\makecell[c]{Number of ``Channels''}&	\makecell[c]{2048} &	\makecell[c]{150} &	\makecell[c]{40 } &	\makecell[c]{60}&	\makecell[c]{40}  &\makecell[c]{30}\\
		\hline
		\makecell[c]{Number of Layers}&5	&5 &3		&5	&5 &5	\\
		\hline
		\makecell[c]{Number of Weights}&	17.5M&	1.66M &959k	&35.9k &22.7k & 15.4k\\
		\hline\hline
	\end{tabular}
	\vspace{-2mm}
\end{table}

\begin{table}[htb!]
\vspace{-2mm}
\setlength\tabcolsep{2pt}
	\centering
	\caption{Sample complexity ($K=3$, $N_s=6$, SNR=10$\rm dB$)}\label{table:samples}
	\vspace{-2mm}
	\footnotesize
	\renewcommand\arraystretch{1}
	\begin{tabular}{c|c|c|c|c|c|c|c}
		\hline\hline
		$N_t$ & FNN & CNN &Vertex-GNN & 1D-GNN & 2D-GNN & 2D'-GNN & 3D-GNN   \\\hline
		8  &480k   & 190k &45k &42k  &20k &35k &15k   \\\hline
		16 & $>$500k& 300k &70k  &30k &18k &26k  &11k  \\\hline
		36 & $>$500k& 310k &100k &25k &15k  &20k &11k  \\\hline
		64 & $>$500k& 420k &135k &40k &22k &33k &14k   \\
		\hline\hline
	\end{tabular}
	\vspace{-2mm}
\end{table}

We also evaluate the inference time under the settings in Table  \ref{table:samples}. Taking $N_t=16$ as an example, the running time of FNN, CNN, vertex-GNN, 1D-, 2D-, {\bf 3D}-GNNs, and the MO algorithm are respectively 1.15, 4.37, 6.63, 1.12, 2.14, {\bf 4.79}, and 7452 milliseconds on CPU.

\begin{figure}[!htb]
	\centering
	\includegraphics[width=0.8\linewidth]{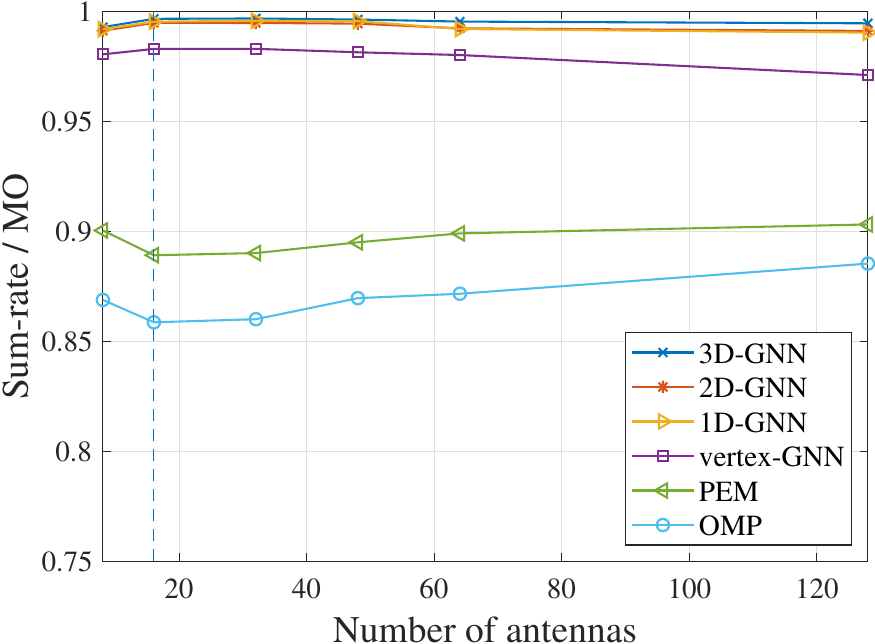}
	\vspace{-2mm}
	\caption{Generalizability to $N_t$, $K=3,N_s=6$.}
	\label{fig:gen-an}
	\vspace{-2mm}
\end{figure}

\begin{figure}[!htb]
	\centering
	\includegraphics[width=0.8\linewidth]{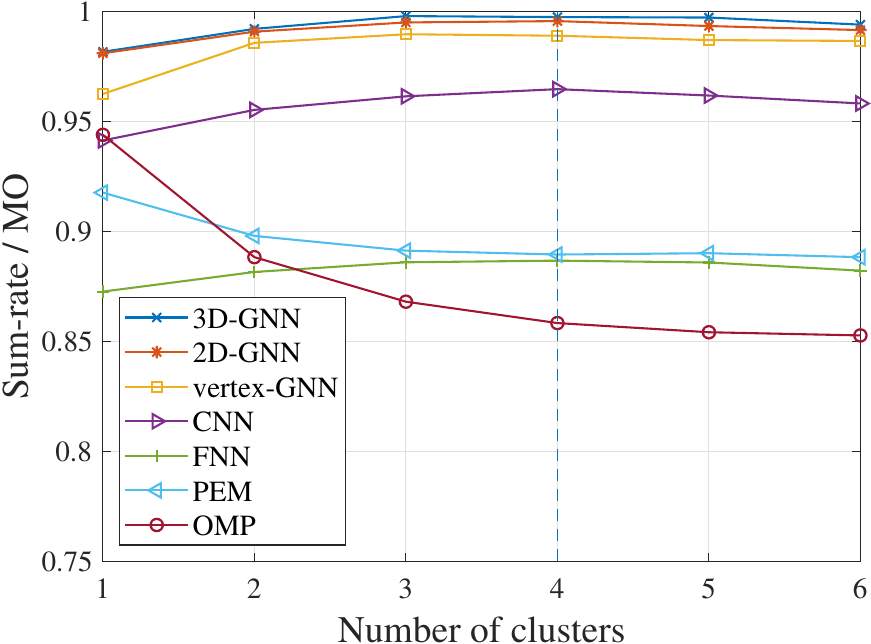}
	\vspace{-2mm}
	\caption{Generalizability to channels, $K=3, N_s=6, N_t=16$.}
	\label{fig:gen-ch}
	\vspace{0mm}
\end{figure}

\begin{figure}[!htb]
	\centering
	\subfigure[SNR = 0 dB]{
		\begin{minipage}[c]{\linewidth}
			\centering
			\includegraphics[width=.8\linewidth]{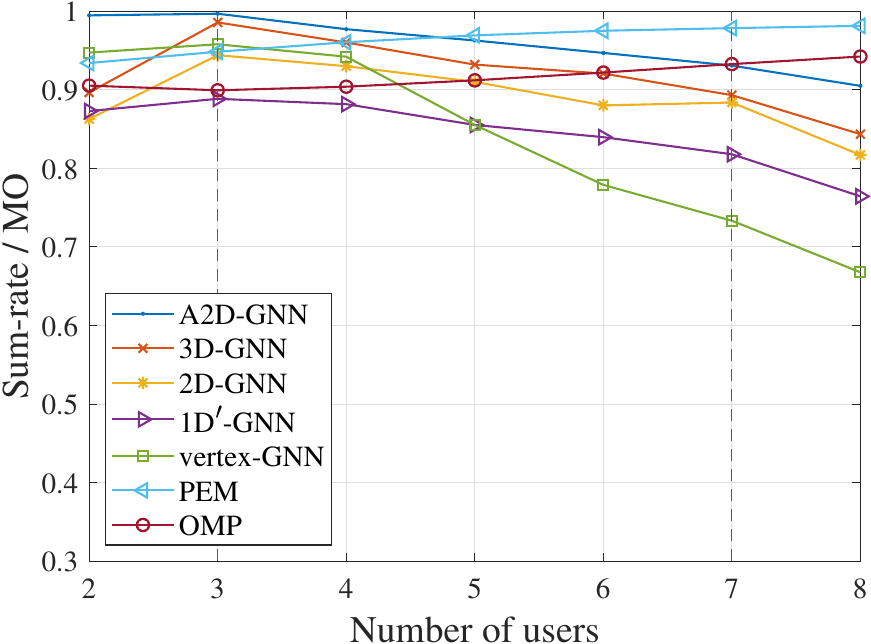}
			\vspace{1mm}
		\end{minipage}
	}\\	 \vspace{-2mm}
	\subfigure[SNR =10 dB]{
		\begin{minipage}[c]{\linewidth}
			\centering
			\includegraphics[width=.8\linewidth]{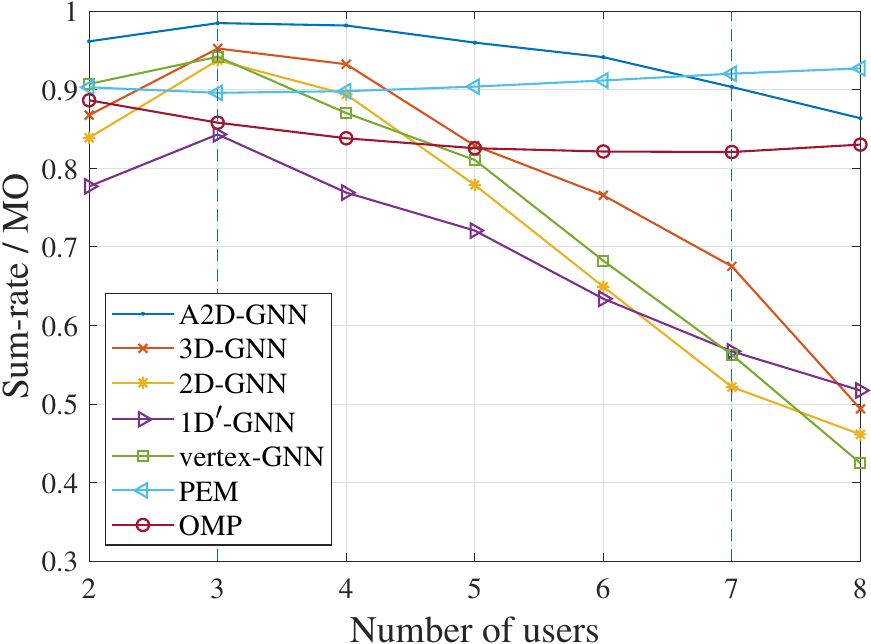}
			\vspace{1mm}
		\end{minipage}
	}\vspace{-2mm}
	\caption{Generalizability to $K$, $N_s=12, N_t=16$.}
	\label{fig:gen-ue}\vspace{-4mm}
\end{figure}

Next, we evaluate the generalization ability of the GNNs to dynamic wireless environments, which are unseen during training.
In Fig. \ref{fig:gen-an}, we show the generalizability of GNNs to the number of antennas, given that FNNs and CNNs cannot be generalized to problem scales. All the GNNs are trained using the samples generated in a system with $N_t=16$ but tested using the samples with $N_t=8 \sim 128$, which all use $\rm mean(\cdot)$ as the pooling function.
%Vertex-GNN has worse performance owing to the information loss.
%fact that the FNN (in users and RF chains) processing the information from each antenna produces a fixed-length vector (64 neurons we set). If there are excessive number of antennas, say 128, the information will be compressed after pooling. However, in MD-GNNs, the updating elements are directly related to all antennas, i.e., antenna-vertices in 1D-GNN, edges between users and antennas in 2D-GNN, hyper-edges connected users, antennas, and RF chains in 3D-GNN. The number of hidden representations grows with the number of antennas, hence each element has the opportunity to be updated individually and information loss is eliminated.
In Fig. \ref{fig:gen-ch}, we show the generalizability of DNNs to channel distribution.
All DNNs are trained using the samples generated by the SV channel model with four scattering clusters, but tested in the channels with the number of clusters varying from one to six, where $N_{ray} = 3$ in each cluster. It can be seen that all GNNs can be well generalized to the number of antennas and all DNNs can be well generalized to the channel distribution.
In Fig. \ref{fig:gen-ue}, we show the generalizability of GNNs to the number of users, where 1D'-GNN only harnesses the permutation of users. All GNNs are trained with samples generated in systems with $K=3$ and 7, but are tested in the systems where the number of users varies from two to eight. The results indicate that A2D-GNN exhibits stronger generalizability, but all GNNs cannot be well-generalized to $K$ at high SNR.

\vspace{4mm}
\subsection{Learning Wideband Hybrid Precoding Policy from Problem $P2$}
A BS with 16 antennas and six RF chains serves three users in a MISO-OFDM system. The SNR defined as $P_{tot}/M{\sigma ^2}$ is 10dB. The channels are generated with the wideband mmWave channel model with uniform
linear array in section \ref{sec:sample}, where $p(\cdot)$ is raised-cosine filter and $N_{cl}=5$, $N_{ray}=10$ \cite{MO}.  Other settings are the same as those in section \ref{sec:sample}.

We simulate the 4D-GNN proposed in section \ref{subsubsec:OFDM}, a 3D-GNN ignoring the permutation of RF chains, and a 2D-GNN further ignoring the permutation of users. We also simulate a CNN, where  channel matrices in different subcarriers are concatenated as a long matrix \cite{learnOFDM}. The results are provided in Table \ref{table:OFDM_M}. We can see that when the frequency selectivity of channels becomes stronger, the sum-rate achieved by the policy learned with CNN reduces  but the stronger permutability of samples improves the learning performance of GNNs. Again, 4D-GNN outperforms the 3D-GNN, and both outperform 2D-GNN.
%We have to admit that the limitations of time and memory of GPU cannot be ignored.
%Although 4D-GNN develop all prior knowledge, it can not be trained with large number of  layers and channels, especially for 16 subcarriers, which leads to relatively poorer performances than other GNNs.

\begin{table}[htb!]
	\centering
	\vspace{-2mm}
	\caption{Sum-rate versus number of Subcarriers ($K=3$, $N_s=6$, $N_t=16$, SNR=10$\rm dB$)}\label{table:OFDM_M}
	\vspace{-2mm}
	\footnotesize
	\renewcommand\arraystretch{1}
	\begin{tabular}{c|c|c|c|c|c|c}	
		\hline\hline
		\multirow{2}{*}{$M$} &\multirow{2}{*}{$D$} & \multirow{2}{*}{\makecell[c]{MO \\ (bps/Hz)}} & \multicolumn{4}{c}{Sum-rate relative to MO}\\
		\cline{4-7}
		 & &  & CNN & 2D-GNN &3D-GNN &4D-GNN \\
		 \hline
		8&	2&	13.62&    86.11\%&  94.95\%&  96.01\%& 96.29\%\\
		\hline
		8&	3&	13.25&  77.36\%&  96.72\%&  97.70\%& 98.08\%\\
		\hline
		16&	2&	13.66& 	85.13\%& 92.91\%& 93.68\%& 93.88\%\\
		\hline
		16&	4&	12.88&    75.51\%&  94.64\%&  95.11\%& 95.40\%\\
		\hline\hline
	\end{tabular}\vspace{-2mm}
\end{table}

{\bf Remark 6:} We have also simulated the 2D-GNN designed in section \ref{subsec:depGNN} for learning the power control policy in \cite{shenyifei}, which is a vertex-level task.
Our results in the setting at SNR = 10 dB show that the 2D-GNN slightly outperforms the vertex-GNN designed in \cite{shenyifei} that uses two FNNs for processing and combining. For example, when there are 50 transceiver pairs, the sum-rate relative to the weighted MMSE algorithm in \cite{WMMSE} achieved by the 2D-GNN is 103.8\% and the relative sum-rate achieved by the vertex-GNN is 102.5\%. Besides, the 2D-GNN achieves better generalizability to the number of transceiver pairs than the vertex-GNN.

\section{Conclusions}\label{sec:conclu}
In this paper, we proposed a unified framework of GNNs to learn wireless policies for avoiding information loss meanwhile exploiting permutation prior. To avoid the information loss, the dimension of the MD-GNN is equal to the number of vertex types. To harness all possible permutations of a problem, the number of vertex types of a graph should be identical to the number of all sets in the problem. To show the potential of GNNs in leveraging permutation prior, we provided a systematic approach to identify sets and model
graphs from optimization problems.
We classified the permutable problems according to the types of sets and the relation between sets. When using the MD-GNNs to learn policies from the problems in different categories, only the input layer, parameter sharing in each update layer, and output layer differ. We mainly took hybrid precoding in mmWave MU-MIMO systems as examples to show how to  construct graphs and design the MD-GNNs. We analyzed the permutability of wideband mmWave channels, which affects the sample complexity of the GNNs for learning precoding policies. We further suggested how to trade off training, inference, and design  complexities by ignoring some permutations deliberately. Simulation results showed that the proposed MD-GNNs outperform the state-of-the-art GNNs, but their time complexities grow with the dimension.

\vspace{-2mm}
\begin{appendices}
\numberwithin{equation}{section}
\section{Proof of Proposition 1}\label{app:A}
\vspace{-2mm}
Without loss of generality, one update layer is considered, where $f_{in}(\cdot)$ satisfies $\overline{\bf X}_1=f_{in}({\bf \Pi}_1^T{\bf H}{\bf \Pi}_2,{\bf \Pi}_1^T{\bm \beta},P_{tot};{\bf \Pi}_{3}^T{\bf a})$, $f_1(\cdot)$ satisfies $\overline{\bf X}_2=f_1(\overline{\bf X}_1)$, and $f_{out}(\cdot)$ satisfies $({\bf \Pi}_2^T{\bf W}_{RF}{\bf \Pi}_{3},{\bf \Pi}_{3}^T{\bf W}_{BB}{\bf \Pi}_1)=f_{out}(\overline{\bf X}_2)$.
Upon substituting $\overline{\bf X}_1$, the composite function $f_1(f_{in}(\cdot))$ satisfies $\overline{\bf X}_2=f_1(f_{in}({\bf \Pi}_1^T{\bf H}{\bf \Pi}_2,{\bf \Pi}_1^T{\bm \beta},P_{tot};{\bf \Pi}_{3}^T{\bf a}))$. By substituting $\overline{\bf X}_2$ into $f_{out}(\cdot)$, we obtain that $({\bf \Pi}_2^T{\bf W}_{RF}{\bf \Pi}_{3},{\bf \Pi}_{3}^T{\bf W}_{BB}{\bf \Pi}_1)=f_{out}(f_1(f_{in}({\bf \Pi}_1^T{\bf H}{\bf \Pi}_2,{\bf \Pi}_1^T{\bm \beta},P_{tot};{\bf \Pi}_{3}^T{\bf a})))$, which satisfies the three-set permutation property.
\end{appendices}

\bibliography{IEEEabrv,myref}

\begin{IEEEbiography}
[{\includegraphics[width=1in,height=1.25in,clip,keepaspectratio]{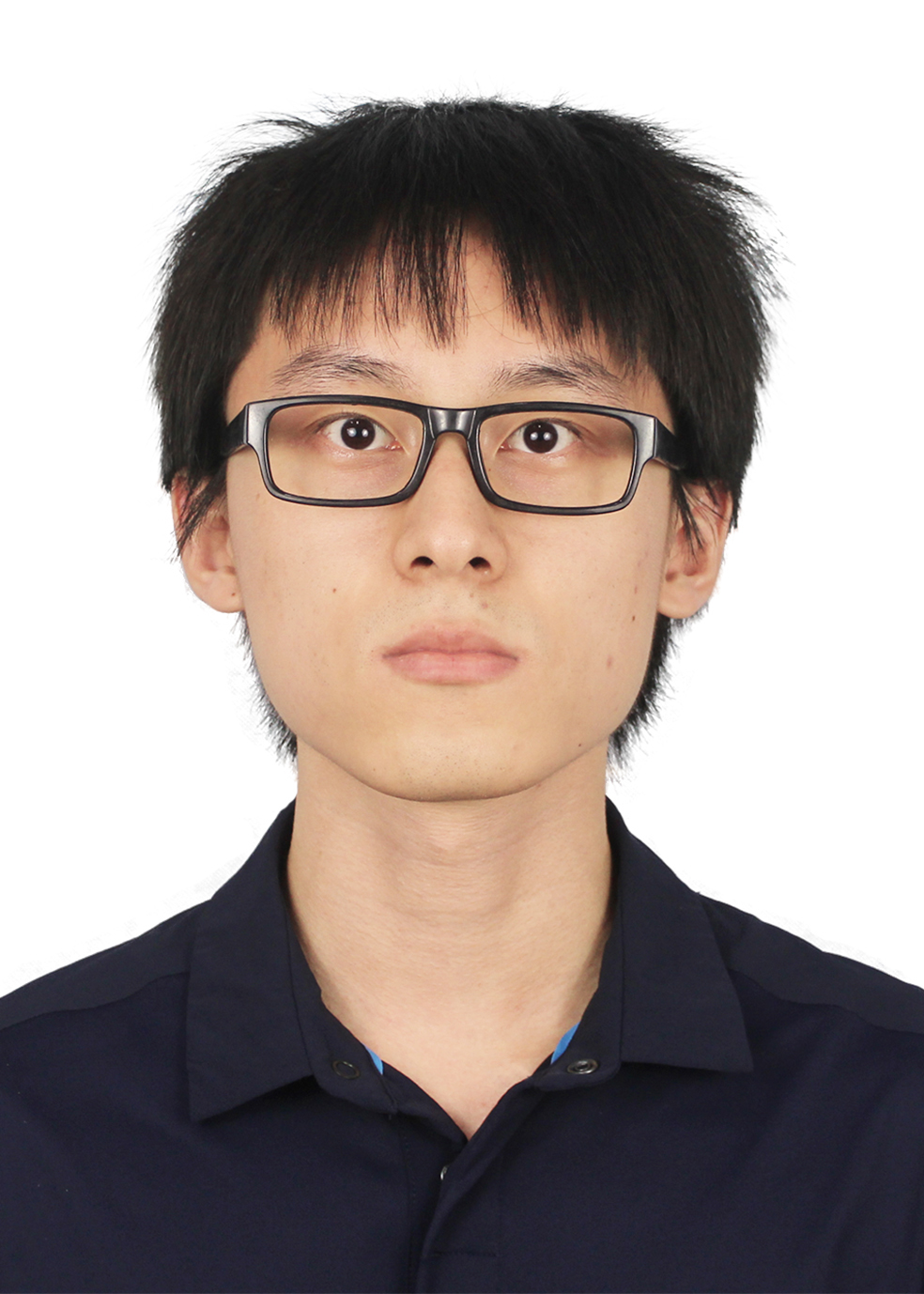}}]{Shengjie Liu}
received his B.S. degree in electronics engineering from Beihang University, China, in 2021. He is currently pursuing his Ph.D. degree in signal and information processing with the School of Electronics and Information Engineering, Beihang University. His research interests include graph neural network and its applications in wireless communications.
\end{IEEEbiography}

\begin{IEEEbiography}
[{\includegraphics[width=1in,height=1.25in,clip,keepaspectratio]{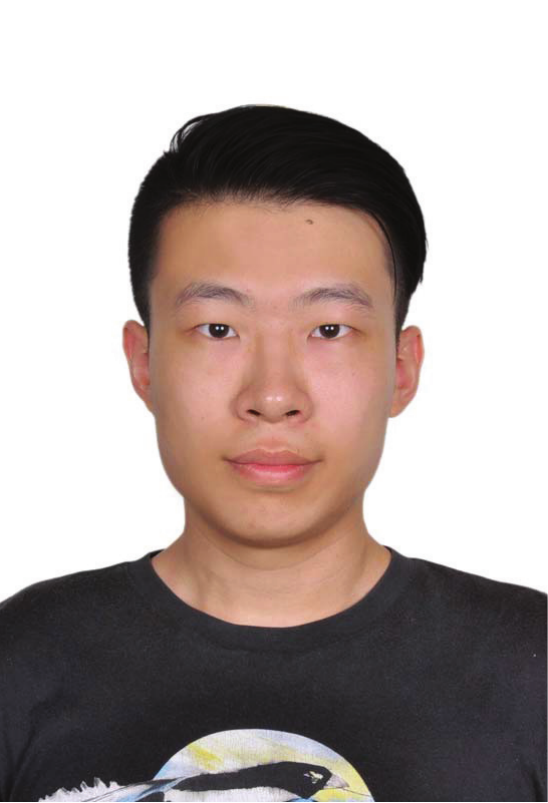}}]{Jia Guo}
(Graduate Student Member, IEEE) received his B.S. degree in electronics engineering and M.S. degree in information and communication engineering from Beihang University, China, in 2016 and 2019, respectively. He is currently pursuing his Ph.D. degree in signal and information processing with the School of Electronics and Information Engineering, Beihang University. His research interests lie in the area of machine learning for wireless communications.
\end{IEEEbiography}

\begin{IEEEbiography}
[{\includegraphics[width=1in,height=1.25in,clip,keepaspectratio]{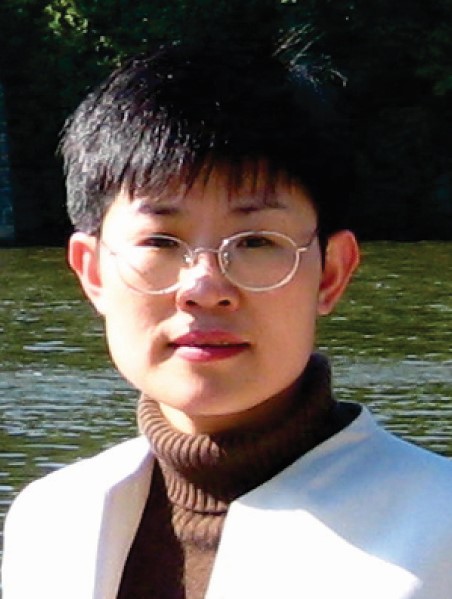}}]{Chenyang Yang}
(Senior Member, IEEE) received the Ph.D. degree in electrical engineering from Beihang University, China, in 1997. She has been a Full Professor with Beihang University since 1999. She has published over 300 articles in the fields of machine learning for wireless communications, URLLC, energy efficient resource allocation, wireless caching, and interference management. She was supported by the first Teaching and Research Award Program for Outstanding Young Teachers of Higher Education Institutions from the Ministry of Education of China. She has served as an associate editor or the guest editor for several IEEE journals. Her recent research interests include mobile/wireless AI, and URLLC.
\end{IEEEbiography}

\end{document}